\documentclass[aps,prx,showkeys,showpacs,twocolumn,longbibliography,superscriptaddress,notitlepage,floatfix]{revtex4-2}
\usepackage{amsmath,amsfonts,amssymb,color,epsfig,graphics,graphicx,latexsym,theorem,url,multirow,diagbox}
\usepackage[colorlinks,colorlinks,citecolor=blue,linkcolor=blue,urlcolor=blue]{hyperref}
\usepackage[utf8]{inputenc}
\usepackage{booktabs}
\usepackage[T1]{fontenc}
\usepackage{courier}
\usepackage{listings}
\usepackage{dcolumn}
\usepackage{comment}
\usepackage{times}

\usepackage{algorithm}  
\usepackage{algorithmicx}  
\usepackage{algpseudocode}  
\usepackage{braket}
\usepackage{bm}

\begin{document}

\title{Recent advances for quantum classifiers}

\begin{abstract}
Machine learning has achieved dramatic success  in a broad spectrum of applications. Its interplay with quantum physics may lead to unprecedented perspectives for both fundamental research and commercial applications, giving rise to an emergent research frontier of quantum machine learning. Along this line, quantum classifiers, which are quantum devices that aim to solve classification problems in  machine learning, have attracted tremendous attention recently. In this review, we give a relatively comprehensive overview for the studies of quantum classifiers, with a focus on recent advances. First, we will review a number of quantum classification algorithms, including quantum support vector machines, quantum kernel methods, quantum decision tree classifiers, quantum nearest neighbor algorithms, and quantum annealing based classifiers. Then, we move on to introduce the variational quantum classifiers, which are essentially variational quantum circuits for classifications. We will review different architectures for constructing variational quantum classifiers and introduce the barren plateau problem, where the training of quantum classifiers might be  hindered by the exponentially vanishing gradient.  In addition, the vulnerability aspect of quantum classifiers in the setting of adversarial learning and the recent experimental progress on different quantum classifiers will also be discussed. 
\end{abstract}

\author{Weikang Li}
\affiliation{Center for Quantum Information, IIIS, Tsinghua University, Beijing 100084, People\textquoteright s Republic of China}
\author{Dong-Ling Deng}\email{dldeng@tsinghua.edu.cn}
\affiliation{Center for Quantum Information, IIIS, Tsinghua University, Beijing 100084, People\textquoteright s Republic of China}
\affiliation{Shanghai Qi Zhi Institute, 41th Floor, AI Tower, No. 701 Yunjin Road, Xuhui District, Shanghai 200232, China}



\maketitle

\tableofcontents

\section{Introduction}\label{introduction}

The interplay between machine learning and quantum physics may lead to unprecedented perspectives for both fields
\cite{Sarma2019Machine}.
In recent years, machine learning has
revolutionized many new techniques in both research and commercial fields, ranging from image recognition to automated driven cars
\cite{Lecun2015Deep,Goodfellow2016Deep,Jordan2015Machine}.
More strikingly, some notoriously challenging problems for traditional methods have been cracked successfully.
Notable examples include playing the game of Go with deep neural networks and tree search \cite{Silver2016Mastering,Silver2017Mastering}, classifying skin cancers with convolutional neural networks \cite{Esteva2017Dermatologistlevel}, and predicting protein structures with the AlphaFold system \cite{Senior2020Improved}.
Proceeding along this direction, machine learning methods may likewise shed new light on solving complex problems in quantum physics \cite{Carleo2019Machine,Dunjko2018Machine,Chng2017Machine,Torlai2018Neuralnetwork,Wang2016Discovering,Luiz2021Machine}.
In parallel, quantum computing has made fascinating progress in recent years, with the experimental demonstrations of quantum supremacy marked as the latest milestone \cite{Arute2019Quantum,Zhong2020Quantum, Wu2021Strong,Zhu2021Quantum}. Nowadays, a number of notable algorithms \cite{Grover1997Quantum,Shor1997PolynomialTime,Childs2010Quantum,Long2001Grover,Harrow2009Quantum,Lloyd2014Quantum,Rebentrost2014Quantum,Dunjko2016QuantumEnhanced,Montanaro2016Quantum,Havlicek2019Supervised,Schuld2019Quantum,Gao2018Quantum,Lloyd2018Quantum,Hu2019Quantum,Huang2021Experimental,Martyn2021Granda}, such as the Harrow-Hassidim-Lloyd algorithm \cite{Harrow2009Quantum}, quantum principal component analysis \cite{Lloyd2014Quantum}, quantum-enhanced feature space \cite{Havlicek2019Supervised,Schuld2019Quantum}, quantum generative models \cite{Gao2018Quantum,Lloyd2018Quantum,Hu2019Quantum,Huang2021Experimental}, quantum support vector machines \cite{Rebentrost2014Quantum}, etc., have been proposed and shown to hold the potential of exhibiting quantum advantages over their classical counterparts. 
These advances have ignited tremendous interest in exploring enhanced machine learning models with quantum devices.

Classifications, as one of the most important branches in machine learning, are nowadays widely applied in commercial and academic applications ranging from face recognition and recommendation systems to earthquake detection and disease diagnosis \cite{Lecun2015Deep,Goodfellow2016Deep,Jordan2015Machine,Esteva2017Dermatologistlevel}.
With the rapid development of quantum computing theory in the past few years, 
it is promising to develop quantum enhanced classification models that may be able to handle complex classification tasks.
To date, a number of quantum classification models and relevant techniques have already been proposed theoretically with some of them even experimentally demonstrated \cite{Bartkiewicz2020Experimental,Peters2021Machine,Haug2021Largescale,Li2015Experimental,Schuld2017Implementing,Blank2020Quantum,Schuld2019Quantum,Rebentrost2014Quantum,Havlicek2019Supervised,Zhu2019Training,Zhao2019Building,Tacchino2019Artificial,Schuld2020Circuit,Schuld2019Evaluating,Mitarai2018Quantum,Lu2020Quantum,Liu2018Differentiable,Li2017Hybrid,Killoran2019Continuous,Grant2018Hierarchical,Farhi2018Classification,Cong2019Quantum,Beer2020Training,Adhikary2020Supervised,Wang2020Quantum,Chen2020Hybrid,Tuerkpence2019Steady,Lloyd2013Quantum,Wiebe2014Quantum,Lu2014Quantum,Heese2021Representation,Mari2021Estimating,Cai2015Entanglement,Stokes2020Quantum,Koczor2019Quantum,Liu1989Limited,Lavrijsen2020Classical,Haug2021Natural,Wierichs2021General,Sweke2020Stochastic,Koczor2020Quantum,Li2021Quantum,Romero2018Strategies,Blance2021Quantum,Kerenidis2019Quantum,Liu2021Hybrid,Wei2021Quantum,Yano2020Efficient,Li2020Quantuma,Guerreschi2017Practical,Kyriienko2021Generalized,Hur2021Quantum,MacCormack2020Branching,LaRose2020Robust,Johri2021Nearest,Jiang2021Quantum,Neven2008Training,Neven2009Training,Neven2009Binary,Carapito2021Identification,Li2018Quantum,Li2021Quantumb,Mott2017Solving,Pudenz2013Quantum,Zlokapa2020Quantum,Nath2021Review,Boyda2017Deploying,Caldeira2020Restricted,Denchev2012Robust,Dixit2021Training,DulnyIII2016Developing,Liu2018Adiabatic,Nguyen2018Image,Willsch2020Support,Herrmann2021Realizing}.
These works have enriched the studies of quantum classifiers from various aspects and may provide valuable guidance for more advanced models in the future \cite{Bharti2021Noisy,Cerezo2020Variational,Preskill2018Quantum}.

In general, the classification process involves a mapping from the input data to different categories, which can also be described as attaching labels to the unlabeled samples.
For supervised learning tasks, it requires that there is already some labeled data serving as the training set, e.g. some handwritten digits with their true labels.
After the learning process, the trained model is expected to master the data's pattern and to be used to classify the unseen data.
In the early studies of quantum classifiers, some works have extended popular classical classification algorithms to the quantum domain.
Renowned examples  include the quantum support vector machine
\cite{Rebentrost2014Quantum}, quantum nearest neighbor algorithms
\cite{Lloyd2013Quantum,Wiebe2014Quantum}, and quantum decision tree classifiers
\cite{Lu2014Quantum,Heese2021Representation}.
Rather than merely implementing the extensions from classical to quantum,
among them there are also approaches exhibiting the potential of providing quadratic or even exponential speedups.
Moreover, it is desirable to design new quantum classifiers by encapsulating the above mentioned algorithms, such as the Harrow-Hassidim-Lloyd algorithm \cite{Harrow2009Quantum} and quantum principal component analysis \cite{Lloyd2014Quantum}, as subroutines.
Recently, quantum kernel methods have drawn a lot of attention.
In Ref. \cite{Rebentrost2014Quantum}, the authors have proposed to use quantum devices to efficiently prepare the linear or polynomial kernel matrices.
When dealing with a classification task, the choice of the input feature space is crucial for the classification performance, e.g. the support vector machine with a linear kernel may not work to classify a dataset that is not linearly separable.
Along this line, several quantum kernel methods have been reported in both theoretical studies and experimental demonstrations \cite{Schuld2017Implementing,Blank2020Quantum,Schuld2019Quantum,Havlicek2019Supervised,Bartkiewicz2020Experimental,Peters2021Machine,Haug2021Largescale,Liu2021Rigorous}.
In particular, Refs. \cite{Schuld2019Quantum,Havlicek2019Supervised,Bartkiewicz2020Experimental,Peters2021Machine,Haug2021Largescale} propose to use the quantum devices to estimate the kernel matrices and leave the rest computations to classical computers,
which might be applicable in the age of noisy intermediate-scale quantum devices
\cite{Preskill2018Quantum}.
It is also worth mentioning that, in Ref. \cite{Liu2021Rigorous}, the authors have proposed a quantum classification algorithm with a rigorous and robust speedup, assuming the hardness of the renowned discrete logarithm problem.

The most successful programming paradigm in machine learning relies on artificial neural networks, a highly abstracted and simplified model of the human brain \cite{Lecun2015Deep,Goodfellow2016Deep}. An artificial neural network consists of a collection of connected units or nodes called artificial neurons, which are typically organized in layers.
There are various types of neural networks, including feedforward \cite{Bebis1994Feedforward,Svozil1997Introduction}, convolutional \cite{Lawrence1997Face}, recurrent \cite{Mikolov2011Extensions,Zaremba2015Recurrent}, and capsule neural networks \cite{Hinton2011Transforming,Hinton2018Matrix,Sabour2017Dynamic,Xi2017Capsule,Xinyi2018Capsule}, each of which bears their own particular structures and levels of complexity.
These networks usually contain a large amount of parameters that can be adjusted during the training procedure.
Due to the well-designed structures and the nonlinear components, e.g. the activation functions and the pooling layers,
some neural networks exhibit outstanding expressive power as classifiers.
Inspired by the achievements in classical machine learning, variational quantum classifiers, which inherit some features of the classical neural networks, have attracted a wide range of attention and rapid progress has been made over the recent years
\cite{Cerezo2020Variational}.
Similar to the classical case, variational quantum circuits contain variational parameters that can be optimized during the training procedure.
During the optimization process,
a key step is to calculate the derivatives of the target functions with respect to the circuit parameters 
\cite{Guerreschi2017Practical,Koczor2020Quantum,Mari2021Estimating,Schuld2020Circuit,Mitarai2018Quantum,Li2017Hybrid,Schuld2019Evaluating,Haug2021Natural,Wierichs2021General,Stokes2020Quantum,Sweke2020Stochastic,Koczor2019Quantum,Lavrijsen2020Classical,Liu1989Limited,Kyriienko2021Generalized}.
In particular, the widely used ``parameter shift rule'' has been proposed and developed
\cite{Mitarai2018Quantum,Li2017Hybrid,Schuld2019Evaluating,Wierichs2021General,Mari2021Estimating}.
In addition to the optimization strategies,
a variety of variational quantum classifiers with different structures have emerged
\cite{Adhikary2020Supervised,Cong2019Quantum,Killoran2019Continuous,Grant2018Hierarchical,Schuld2020Circuit,Farhi2018Classification,Blance2021Quantum,Kerenidis2019Quantum,Liu2021Hybrid,Wei2021Quantum,Yano2020Efficient,Li2020Quantuma,Hur2021Quantum,MacCormack2020Branching}.
Among these works, some of them extend straightforwardly the concepts from the classical neural networks to the quantum domain, e.g. the quantum convolutional neural networks \cite{Cong2019Quantum} and the continuous-variable quantum neural networks (which provide the frameworks suitable for constructing convolutional networks, recurrent networks \textit{et al}.) \cite{Killoran2019Continuous}.
Some other works use special structures inspired by physics to build quantum classifiers, e.g. the tree tensor network classifiers, the multi-scale entanglement renormalization ansatz classifiers
\cite{Grant2018Hierarchical} and the multi-level quantum system classifiers
\cite{Adhikary2020Supervised}. 
It is worth mentioning that there are algorithms that search for appropriate structures of the variational quantum circuits for certain tasks \cite{Rattew2019Domain,Li2017Approximate,Zhang2020Differentiable,Cincio2018Learning,Cincio2021Machine,Lu2020Markovian,Chivilikhin2020Mog,Pirhooshyaran2020Quantum,Ostaszewski2021Structure,Li2020Quantum,Foesel2018Reinforcement,Bolens2020Reinforcement,Wang2021QuantumNAS,Bilkis2021Semiagnostic,Du2020QuantumCircuit,Grimsley2019Adaptive,Tang2021QubitADAPTVQE}, e.g. a quantum neuroevolution algorithm that autonomously finds suitable quantum neural networks \cite{Lu2020Markovian} and a differentiable quantum architecture search algorithm that allows automated quantum circuit designs in an end-to-end differentiable fashion \cite{Zhang2020Differentiable}.

\begin{figure*}
    \centering
    \includegraphics[width=1\textwidth]{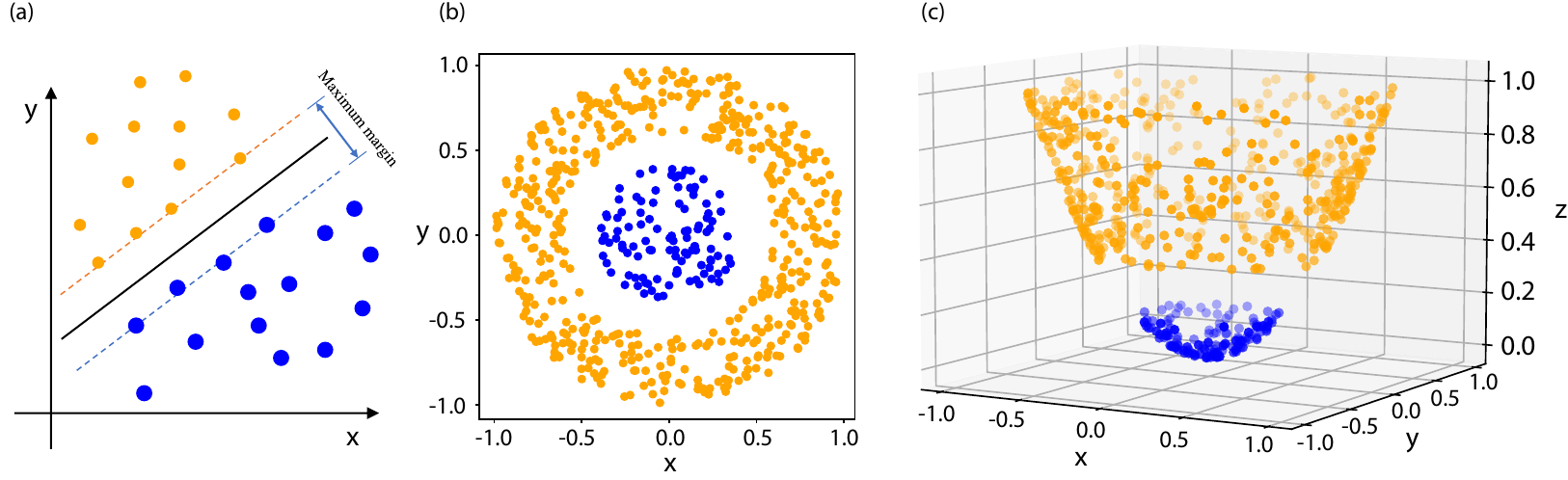}
    \caption{(a) A schematic diagram for support vector machines classifying a dataset, where the two kinds of data points (marked as orange and blue balls, respectively) are linearly separable. (b) A situation where the dataset is not linearly separable. (c) A method to solve the problem in (b) using kernel methods to map the data to a higher dimensional space.}
    \label{Fig:SVM}
\end{figure*}

Though quantum classifiers have drawn a wide range of interest and investigation, there are crucial challenges that may hinder their performance and should be studied in depth: the barren plateau phenomena during the training process and vulnerability of quantum classifiers.
First, the barren plateaus refer to the phenomenon in the optimization procedure where the gradients concentrate to zero exponentially fast as the system size increases \cite{Shalev-Shwartz2017Failures}.
To illustrate the phenomenon, we consider the initial parameters in the high-dimensional parameter space with the cost function being an additional dimension.
At ideal conditions, during the training process the parameters can converge to the point corresponding to the minimal value of the cost function with decent speed,
just like a snow ball rolling down the hill.
However, the barren plateaus present the situation that the cost function with respect to the parameters is too flat in most of the space except a small area around the optimal point,
thus it is nearly impossible to converge to the optimal point.
Nowadays, this phenomenon has attracted broad interest across communities, with many works analyzing the reasons why barren plateau occurs and designing strategies to avoid it
\cite{Mcclean2018Barren,Cerezo2021Cost,Pesah2020Absence,Wang2020Noise,Huembeli2021Characterizing,
Patti2021Entanglement,Marrero2020Entanglement,Uvarov2021Barren,Arrasmith2020Effect,Wierichs2020Avoiding,
Cerezo2021Higher,Grant2019Initialization,Haug2021Optimal,Holmes2021Barren,Holmes2021Connecting,
Zhao2021Analyzing,Kieferova2021Quantum,Liu2021Presence,Skolik2020Layerwise}.
The reasons underlying this phenomenon include the poor design of the quantum circuits \cite{Mcclean2018Barren},  non-negligible noises \cite{Wang2020Noise},  massive entanglement \cite{Patti2021Entanglement}, and inappropriate choice of cost functions \cite{Cerezo2021Cost, Liu2021Presence}. 
Second, the vulnerability of quantum classifiers concerns their performance against adversarial attacks.
Adversarial examples refer to input samples that an attacker has crafted in order to fool the machine learning models. 
For a simple example, a classifier is supposed to assign a picture to the class ``dog''. If we add a carefully designed and imperceptible perturbation to this picture, then it may be assigned to the class ``cat'' incorrectly by the classifier.
The vulnerability of quantum machine learning models to crafted adversarial examples as well as the design of defense strategies are actively investigated at the current stage, sparking an emergent research direction of quantum adversarial machine learning  \cite{Lu2020Quantum,Liu2020Vulnerability,Liao2021Robust,Weber2021Optimal,Du2021Quantum,Gong2021Universal}.

In parallel to the theoretical development, stimulating progress has also been made in the experimental side
 \cite{Li2015Experimental,Schuld2017Implementing,Blank2020Quantum,Havlicek2019Supervised,Bartkiewicz2020Experimental,Cai2015Entanglement,Johri2021Nearest,Peters2021Machine,Haug2021Largescale}.
For example, short after the theoretical proposal of the quantum support vector machines and quantum nearest neighbor algorithms, in Refs. \cite{Li2015Experimental,Cai2015Entanglement,Johri2021Nearest} the authors have successfully demonstrated the proof-of-principle experiments.
With the commercially available quantum annealers from D-Wave Systems \cite{Johnson2011Quantum,Jones2013Google,Lanting2014Entanglement,Venegas-Andraca2018Crossdisciplinary}, the quantum annealing based classifiers have also been developed and applied to a variety of tasks.
More recently, both the variational quantum classifier \cite{Havlicek2019Supervised} and the estimation of kernel matrices \cite{Bartkiewicz2020Experimental,Havlicek2019Supervised,Peters2021Machine,Haug2021Largescale} have been experimentally implemented.
These quantum experiments may be carried out with higher accuracy and at a lower cost with the development of quantum devices.
And by that time, more practical quantum classifiers might be implemented, and it is appealing to see some of them served in commercial applications.

This review is organized as follows. 
In Section $2$, we will introduce some important algorithms and models for quantum classifiers. This section can be divided into several subsections including the quantum support vector machines, quantum decision tree, quantum nearest neighbors algorithms, quantum annealing based classifiers, and variational quantum classifiers. To better present these advances, we will introduce some relevant classical concepts first and then focus on the quantum models. In addition to the theoretical progress, the experimental advances will be reviewed in Section $3$.
In the last section, we will summarize this review and give an outlook for the future research.

\section{Quantum classifiers}\label{classifier}
\subsection{Quantum support vector machine and kernel methods}\label{qsvm}

Support vector machines (SVMs) are widely used machine learning models and can be applied to find the optimal hyperplane for classification tasks
\cite{Cortes1995Support,Noble2006What,Suthaharan2016Support}.
For a linearly separable dataset given as
\begin{eqnarray}
    \mathcal{D} = \{(\vec{x}_1,y_1),(\vec{x}_2,y_2),\dots,(\vec{x}_M,y_M)\},
\end{eqnarray}
where $M$ is the size of the training dataset with data dimension $N$,
it is feasible to construct an $(N-1)$-dimensional hyperplane with the maximum margin to optimally classify the data in the $N$-dimensional hyperspace. For simplicity, we only consider the binary classification task as an example and denote the labels as $y_i=\pm 1$. 
This basic idea is briefly illustrated in Fig.~\ref{Fig:SVM}(a).

We use a set of parameters $(\vec{\omega},b)$ to denote  a hyperplane in the $N$-dimensional space. In general, the hyperplane that classifies the dataset may not be unique, 
i.e., there exists more than one solution for $\vec{w}$ and $b$ such that for every training data, $y_i(\vec{w} \cdot \vec{x}_i + b) \geq 1$ is satisfied. 
For SVMs, the task is to find the maximum separation between the two classes, i.e., to maximize the distance $\frac{2}{|\vec{w}|}$ between the space $\vec{w} \cdot \vec{x}_i + b \geq 1$ ($+1$ class) and $\vec{w} \cdot \vec{x}_i + b \leq -1$ ($-1$ class). Then, the decision function can be written as
\begin{eqnarray}
f(\vec{x}) = \operatorname{sgn}\left(\vec{w}^* \cdot \vec{x} + b^*\right),\label{Decision function}
\end{eqnarray}
where $\vec{w^*}$ and $b^*$ are the optimal values for the parameters obtained from the optimization procedure.

To solve the original optimization problem (also called the primal problem in the literature), it is helpful to transform it to the corresponding dual problem. With  Lagrange multipliers, the goal can be reduced to maximizing
\begin{eqnarray}
L(\vec{\alpha})=\sum_{j=1}^{M} y_{j} \alpha_{j}-\frac{1}{2} \sum_{j, k=1}^{M} \alpha_{j} (\vec{x}_j \cdot \vec{x}_k) \alpha_{k},
\end{eqnarray}
where \(\sum_{j=1}^{M} \alpha_{j}=0\) and \(y_{j} \alpha_{j} \geq 0\).
Then the weights and bias can be recovered as $\vec{w}^* = \sum_{j=1}^{M} \alpha_{j}^* \vec{x}_j$ and $b^* = y_k - \vec{w}^* \cdot \vec{x}_k$ for an index $k$ corresponding to a non-zero $\alpha_k^*$. Actually, we can directly plug these expressions into the decision function in Eq. (\ref{Decision function}) and obtain
\begin{eqnarray}
f(\vec{x})=\operatorname{sgn}\left(\sum_{j=1}^{M} \alpha_{j}^* (\vec{x}_j \cdot \vec{x})+b^*\right).
\end{eqnarray}

For cases where the dataset is nearly linearly separable, i.e., there are a few outliers that violate the linearity, the hard-margin condition can be relaxed to a soft-margin one: $y_i(\vec{w} \cdot \vec{x}_i + b) \geq 1 - \xi_i$, where $\xi_i$ is a cost \cite{Cortes1995Support,Noble2006What,Suthaharan2016Support}.  
Then we can use the similar optimization procedure to solve the problem.  Since noisy data is more general in the physical world, the soft-margin condition is more often used than the hard-margin condition in practice.

However, 
when the dataset is not linearly separable, such as the example shown in Fig.\ref{Fig:SVM}(b), the above SVM approach may not work. To handle this issue, 
kernel methods are applied to help build a hyperplane in a higher dimensional kernel space and make the data linearly separable again
\cite{Boser1992Training}.
This can be briefly illustrated as follows. 
As shown in Fig.\ref{Fig:SVM}(b), the data samples in the feature space $(x_1,x_2)$  are not linearly separable. 
However, if the feature space can be transformed into $(x_1,x_2,x_1^2+x_2^2)$, 
then it is obvious that the samples in the new basis become linearly separable as shown in Fig.\ref{Fig:SVM}(c). The kernel function is defined as
\begin{eqnarray}
K(x,z) = \phi(x) \cdot \phi(z),
\end{eqnarray}
where $\phi(x)$ is the function mapping the feature space to the kernel space. 
With an appropriate choice of the kernel function, 
the data sample can be mapped into a higher dimensional space and becomes linearly separable. 
We note that $\phi(x)$ does not need to be explicitly defined, 
since we will only use the kernel function to compute the inner product of the vectors in the kernel space. In this case, the dual problem becomes maximizing
\begin{eqnarray}
L(\vec{\alpha})=\sum_{j=1}^{M} y_{j} \alpha_{j}-\frac{1}{2} \sum_{j, k=1}^{M} \alpha_{j} K_{j k} \alpha_{k},
\end{eqnarray}
where \(\sum_{j=1}^{M} \alpha_{j}=0\) and \(y_{j} \alpha_{j} \geq 0\).
After obtaining the optimal parameters for $\vec{\alpha}$, the decision function can then be formulated similar to the linear case as
\begin{eqnarray}
f(\vec{x})=\operatorname{sgn}\left(\sum_{j=1}^{M} \alpha_{j}^* K\left(\vec{x}_{j}, \vec{x}\right)+b^*\right).
\end{eqnarray}
In this way, the kernel methods can be utilized to efficiently handle some non-linear problems in the SVM framework.

We now introduce the quantum support vector machines (QSVMs)
\cite{Rebentrost2014Quantum}, which are based on a least-squares version of classical SVMs 
\cite{Suykens1999Least}.
Compared with the traditional SVMs we introduced above, the least-squares SVMs consider equality type constraints rather than inequality type constraints with slack variables $e_j$ introduced to transform the constraints as
\begin{eqnarray}
y_j(\vec{w} \cdot \vec{x}_j + b) \geq 1 
\Rightarrow 
\vec{w} \cdot \vec{x}_j + b = y_j - y_je_j.
\end{eqnarray}
Besides, a regularization term $\frac{\gamma}{2}\Sigma_{j=1}^M e_j^2$ is added to the Lagrangian to be optimized, where $\gamma$ is a hyperparameter describing the ratio of the Lagrangian's components. As a consequence, the problem after applying the optimization of the Lagrangian can be formulated as a linear equation:
\begin{eqnarray}
\mathbf{F}\left(\begin{array}{l}
b \\
\vec{\alpha}
\end{array}\right)=\left(\begin{array}{cc}
\mathbf{0} & \vec{1}^{T} \\
\vec{1} & \mathbf{K}+\gamma^{-1} \mathbf{I}
\end{array}\right)\left(\begin{array}{l}
b \\
\vec{\alpha}
\end{array}\right)=\left(\begin{array}{l}
0 \\
\vec{y}
\end{array}\right), \label{LinearEqSVM}
\end{eqnarray}
where $\mathbf{K}$ is an $M \times M$ kernel matrix and the matrix $\mathbf{F}$ is $(M+1) \times (M+1)$ dimensional. In the quantum case, the above equation can be rewritten compactly as $\mathbf{\hat{F}} \ket{b,\vec{\alpha}} = \ket{\vec{y}}$, where $\mathbf{\hat{F}} = \mathbf{F}/\operatorname{Tr}(\mathbf{F})$.

With this formulation,  one may consider the  Harrow-Hassidim-Lloyd (HHL) algorithm to solve  the above linear equations \cite{Harrow2009Quantum}. For this purpose, it is necessary that $\mathbf{\hat{F}}$ can be both created and exponentiated efficiently.
For simplicity, $\mathbf{\hat{F}}$ can be written as $(\gamma^{-1} \mathbf{I}+\mathbf{J+K})/\operatorname{Tr}(\mathbf{F})$ with
$\mathbf{J} = \left(\begin{array}{cc}
    \mathbf{0} & \vec{1}^T \\
    \vec{1} & \mathbf{0}
    \end{array}\right).
    $
Actually, with the help of Lie product formula, $e^{-i\mathbf{\hat{F}} \Delta t}$ can be decomposed as
\begin{eqnarray}
e^{-i\mathbf{\hat{F}} \Delta t} = e^{-\frac{i \gamma^{-1} \mathbf{I} \Delta t}{\operatorname{Tr}(\mathbf{F})}}e^{-\frac{i \mathbf{J} \Delta t}{\operatorname{Tr}(\mathbf{F})}}e^{-\frac{i \mathbf{K} \Delta t}{\operatorname{Tr}(\mathbf{F})}} + O(\Delta t^2).
\end{eqnarray}
The exponentiation operations for $\mathbf{I}$ and $\mathbf{J}$ are straightforward, as the eigenvalues and eigenstates for these two matrices can be analytically calculated. 
The main challenges are to create and exponentiate $\mathbf{K}$ efficiently. 
To overcome these challenges, the first step is to call the training data oracle \cite{Lloyd2014Quantum}:
\begin{eqnarray}
1 / \sqrt{M} \sum_{i=1}^{M} \ket{i}
\Rightarrow 
1 / \sqrt{N_{\chi }} \sum_{i=1}^{M}\left|\vec{x}_{i} \| i\right\rangle\left|\vec{x}_{i}\right\rangle,
\end{eqnarray}
where $N_{\chi}=\sum_{i=1}^{M}\left|\vec{x}_{i}\right|^{2}$ denotes the normalization factor.
Then we trace out the second register and obtain the reduced density matrix $\mathbf{\hat{K}} = \mathbf{K}/\operatorname{Tr}(\mathbf{K})$, which accomplishes the challenge of creating $\mathbf{K}$.
For the exponentiation of $\mathbf{\hat{K}}$, Lloyd, Mohseni, and Rebentrost have introduced a powerful method in Ref.
\cite{Lloyd2014Quantum}.
Suppose we want to apply $e^{-i\mathbf{\hat{K}} \Delta t}$ to any density matrix $\rho$. 
Instead of trying to construct $e^{-i\mathbf{\hat{K}} \Delta t}$ directly, the following fact can be utilized
\begin{eqnarray}
\begin{aligned}
    e^{-i\mathbf{\hat{K}} \Delta t} \rho e^{i\mathbf{\hat{K}} \Delta t} & \approx \operatorname{Tr}_{1}\left\{e^{-i S \Delta t} \mathbf{\hat{K}} \otimes \rho e^{i S \Delta t}\right\} \\
&=\rho-i \Delta t[\mathbf{\hat{K}}, \rho]+O\left(\Delta t^{2}\right),
\end{aligned}
\end{eqnarray}
where $\operatorname{Tr_1}$ means tracing out the first register, and $S$ is a SWAP operator which is sparse and easy to exponentiate. This result indicates that we can approximately evolve the density matrix $\rho$ by time $\Delta t$ efficiently using $S$ as the Hamiltonian and tracing out the first register. Back to the task of $\mathbf{\hat{F}}$'s exponentiation,  $\mathbf{\hat{K}}$ and $\mathbf{K}/\operatorname{Tr}(\mathbf{F})$ only differ by a constant, which can be easily fixed via multiplying the time by $\operatorname{Tr}(\mathbf{K})/\operatorname{Tr}(\mathbf{F})$.

After these preparations, it is now straightforward to see that the HHL algorithm can be utilized to solve the linear Eq. (\ref{LinearEqSVM}). 
With the phase estimation algorithm,
the eigenvalues of $\mathbf{\hat{F}}$ can be extracted to the ancillary register. Then with a controlled rotation and the uncomputing strategy, one arrives at the target state $\ket{b,\vec{\alpha}}$. This process can be summarized as
\begin{eqnarray}
|\vec{y}\rangle|0\rangle \rightarrow \sum_{j=1}^{M+1}\left\langle u_{j} \mid \vec{y}\right\rangle\left|u_{j}\right\rangle\left|\lambda_{j}\right\rangle \rightarrow \sum_{j=1}^{M+1} \frac{\left\langle u_{j} \mid \vec{y}\right\rangle}{\lambda_{j}}\left|u_{j}\right\rangle, \quad\;
\end{eqnarray}
where $\ket{u_j}$ represents the eigenstate of $\mathbf{\hat{F}}$ with eigenvalue $\lambda_j$, and the final state is the target state: $\ket{b,\vec{\alpha}}= \frac{1}{\sqrt{C}}\left(b|0\rangle+\sum_{k=1}^{M} \alpha_{k}|k\rangle\right)=\sum_{j=1}^{M+1} \frac{\left\langle u_{j} \mid \vec{y}\right\rangle}{\lambda_{j}}\left|u_{j}\right\rangle$.

With the above procedure, we arrive at a quantum state which encodes all the optimal parameters for the corresponding SVM. 
For  a classification task, the training data oracle can be called again to get $\ket{\vec{u}}$ using $\ket{b,\vec{\alpha}}$ and construct the query state $\ket{\vec{x}}$:
\begin{eqnarray}
|\vec{u}\rangle &=&\frac{1}{\sqrt{N_{\vec{u}}}}\left(b|0\rangle|0\rangle+\sum_{k=1}^{M} \alpha_{k}\left|\vec{x}_{k}\right||k\rangle\left|\vec{x}_{k}\right\rangle\right), \\
|\vec{x}\rangle &=& \frac{1}{\sqrt{N_{\vec{x}}}}\left(|0\rangle|0\rangle+\sum_{k=1}^{M}|\vec{x}||k\rangle|\vec{x}\rangle\right),
\end{eqnarray}
where $N_{\vec{u}}=b^{2}+\sum_{k=1}^{M} \alpha_{k}^{2}\left|\vec{x}_{k}\right|^{2}$ and $N_{\vec{x}}=M|\vec{x}|^{2}+1$ are normalization factors.
To classify $\ket{x}$, intuitively we can calculate the inner product of these two states since we have
\begin{eqnarray}
\langle\vec{u} \mid \vec{x}\rangle=\frac{1}{\sqrt{N_{\vec{x}} N_{\vec{u}}}}\left(b+\sum_{k=1}^{M} \alpha_{k}\left|\vec{x}_{k} \| \vec{x}\right|\left\langle\vec{x}_{k} \mid \vec{x}\right\rangle\right),
\end{eqnarray}
which exactly corresponds to the decision function.
For the physical implementation of the inner product, a SWAP test
\cite{Buhrman2001Quantum} 
can be utilized to project the ancillary qubit in state $1/\sqrt{2}(\ket{0}\ket{\vec{u}}+\ket{1}\ket{\vec{x}})$ to state $1/\sqrt{2}(\ket{0}-\ket{1})$.
Then the success probability is $P = 1/2(1-\langle\vec{u} \mid \vec{x}\rangle)$, which can be used to determine $\ket{x}$'s label according to the sign of $1/2-P$.

So far we have discussed the case of linearly separable dataset for quantum support vector machines. The extension to the nonlinear case is straightforward: one only needs to map each vector for the data sample $\ket{\vec{x}_k}$ into its $d$-times tensor product $\ket{\vec{x}_k}\otimes\cdots\otimes\ket{\vec{x}_k}$ to obtain a polynomial kernel with degree $d$ \cite{Rebentrost2014Quantum}. We mention that this trick is universally applicable, namely it can be used to construct arbitrary polynomial kernels. From the computational cost perspective, it is straightforward to obtain that the time complexity for classical SVMs scales as $O(\log(\frac{1}{\epsilon})\operatorname{poly}(N,M))$, where $\epsilon$ characterizes the desired accuracy. In sharp contrast, the quantum support vector machines  introduced above feature a time complexity of $O(\operatorname{log}(NM))$ for both training and inference, giving rise to an exponential speedup over their classical counterparts. However, it is worthwhile to mention that there indeed exist several possible caveats as discussed in depth by Aaronson \cite{Aaronson2015Read}.

From the above discussion on quantum support vector machines, we see that kernel methods are very useful in certain classification tasks. More recently, a number of different quantum classification models with kernel methods have also been introduced \cite{Schuld2017Implementing,Blank2020Quantum,Schuld2019Quantum,Havlicek2019Supervised,Bartkiewicz2020Experimental}. For instance, in Ref. \cite{Schuld2017Implementing} a distance-based classifier has been proposed with the decision function
\begin{eqnarray}
\tilde{y}=\operatorname{sgn}\left(\sum_{m=1}^{M} y^{m}\left[1-\frac{1}{4 M}\left|\tilde{\mathbf{x}}-\mathbf{x}^{m}\right|^{2}\right]\right),
\end{eqnarray}
where $\tilde{\mathbf{x}}$ is the input to be predicted and $M$ is the size of the training set. The component $1-\frac{1}{4 M}\left|\tilde{\mathbf{x}}-\mathbf{x}^{m}\right|^{2}$ here can be treated as a kernel. For a balanced training set, i.e. the number of samples belonging to the two different classes is equal, the decision function can be reduced to
\begin{eqnarray}
\tilde{y}=\operatorname{sgn}\left(\sum_{m=1}^{M} \left[-\frac{y^{m}}{4 M}\left|\tilde{\mathbf{x}}-\mathbf{x}^{m}\right|^{2}\right]\right).
\end{eqnarray}
Then it is clear that if $\tilde{\mathbf{x}}$ belongs to the $+1$ class, the distance will be small between $\tilde{\mathbf{x}}$ and the samples in the $+1$ class, and large between $\tilde{\mathbf{x}}$ and the samples in the $-1$ class. As a result, the sign function can be used to  assign the input sample to the right class with high probability.

To obtain a quantum analogue of the above algorithm,  the amplitude encoding scheme can be used and the data with $2^n$ features can be embedded into a quantum state with only $n$ qubits.  In this way, the quantum state can be written as $\left|\psi_{\mathbf{x}}\right\rangle=\sum_{i=0}^{N-1} x_{i}|i\rangle$ up to an irrelevant normalization factor. With some data preparation methods such as quantum random access memory (QRAM)
\cite{Giovannetti2008Quantum},
the initial state can be prepared to be
\begin{eqnarray}
|\mathcal{D}\rangle=\frac{1}{\sqrt{2 M}} \sum_{m=1}^{M}|m\rangle\left(|0\rangle\left|\psi_{\tilde{\mathbf{x}}}\right\rangle+|1\rangle\left|\psi_{\mathbf{x}^{m}}\right\rangle\right)\left|y^{m}\right\rangle,
\end{eqnarray}
where $\psi_{\tilde{\mathbf{x}}}$ represents the input data and the index $m$ represents the labeled data.
Then the key step is to apply a Hadamard gate on the ancillary qubit to transform the state to
\begin{eqnarray}
\frac{1}{2 \sqrt{M}} \sum_{m=1}^{M}|m\rangle\left(|0\rangle\left|\psi_{\tilde{\mathbf{x}}+\mathbf{x}^{m}}\right\rangle+|1\rangle\left|\psi_{\tilde{\mathbf{x}}-\mathbf{x}^{m}}\right\rangle\right)\left|y^{m}\right\rangle,
\end{eqnarray}
where $\left|\psi_{\tilde{\mathbf{x}} \pm \mathbf{x}^{m}}\right\rangle=\left|\psi_{\tilde{\mathbf{x}}}\right\rangle \pm\left|\psi_{\mathbf{x}^{m}}\right\rangle$. If a conditional measurement on the ancillary qubit is used to project it into the $\ket{0}\bra{0}$ subspace, it will succeed with probability $\mathrm{p_{acc}}=\frac{1}{4 M} \sum_{m}\left|\tilde{\mathbf{x}}+\mathbf{x}^{m}\right|^{2}$ and the state will collapse into 
\begin{eqnarray}
\frac{1}{2 \sqrt{M \mathrm{p}_{\mathrm{acc}}}} \sum_{m=1}^{M} \sum_{i=1}^{N}|m\rangle\left(\tilde{x}_{i}+x_{i}^{m}\right)|i\rangle\left|y^{m}\right\rangle.
\end{eqnarray}

In the last step, a measurement will be applied on the label qubit $\ket{y^m}$. 
For different outcomes, the corresponding probability will be 
\begin{eqnarray}
\mathrm{p}(\tilde{y}=k)=\frac{1}{4 M \mathrm{p_{acc}}} \sum_{m \mid y^{m}=k}\left|\tilde{\mathrm{x}}+\mathrm{x}^{m}\right|^{2}.
\end{eqnarray}
Then it is straightforward to assign the input sample to the class with higher probability.

Actually, after applying the Hadamard gate, the next operations can be flexible. The post-selection can also be used to choose the state $\ket{1}$ while keeping the rest steps unchanged. In doing so, the probability  can be expressed as
\begin{eqnarray}
\mathrm{p}(\tilde{y}=k)=\frac{1}{4 M \mathrm{p_{acc}^{\prime}}} \sum_{m \mid y^{m}=k}\left|\tilde{\mathrm{x}}-\mathrm{x}^{m}\right|^{2},
\end{eqnarray}
which also corresponds to the decision function above. 

It should be noted that the two different measurement outcomes of the ancillary qubit can both serve as a classifier, while the post-selection scheme will abandon one of them. This idea is reviewed and discussed in Ref. \cite{Blank2020Quantum} with a classification protocol proposed without post-selection. In this work, instead of choosing a particular value of the ancillary qubit to handle the task, a joint measurement is used to measure the expectation value $\left\langle\sigma_{z}^{(a)} \sigma_{z}^{(l)}\right\rangle$ of the ancillary qubit and label qubit. In this way, both measurement outcomes of the ancillary qubit will be covered and the expectation value is 
\begin{eqnarray}
\left\langle\sigma_{z}^{(a)} \sigma_{z}^{(l)}\right\rangle =\sum_{m=1}^{M}(-1)^{y_{m}} \operatorname{Re}(\left\langle\tilde{\mathbf{x}} \mid \mathbf{x}_{m}\right\rangle).
\end{eqnarray}
The decision function can then be constructed by using the sign function.
In fact, the classification scheme introduced in 
\cite{Schuld2017Implementing} and its variant discussed in
\cite{Blank2020Quantum} only use the real part of the quantum state overlap.
To further enrich this classification model,
a SWAP test classifier is also proposed in
\cite{Blank2020Quantum}.
Assume that the initial state is prepared to be
\begin{eqnarray}
\sum_{m=1}^{M} \sqrt{w_{m}}|0\rangle|\tilde{\mathbf{x}}\rangle^{\otimes n}\left|\mathbf{x}_{m}\right\rangle^{\otimes n}\left|y_{m}\right\rangle|m\rangle,
\end{eqnarray}
where $\omega_m$ is a controllable non-negative parameter.
The basic idea is to apply a ``$\mathrm{H}_{a} \cdot \mathrm{C}-\mathrm{SWAP}^{n} \cdot \mathrm{H}_{a}$'' gate (a Hadamard gate on the ancillary qubit, a Controlled-SWAP gate on the input qubits and the training data qubits controlled by the ancillary qubit, and a Hadamard gate on the ancillary qubit again) to the initial state and transform it to
\begin{eqnarray}
\ket{\Psi_f^s} = \sum_{m=1}^{M} \frac{\sqrt{\omega_{m}}}{2}\left(|0\rangle\left|\psi_{n+}\right\rangle+|1\rangle\left|\psi_{n-}\right\rangle\right)\left|y_{m}\right\rangle|m\rangle,\quad
\end{eqnarray}
where $\left|\psi_{n \pm}\right\rangle=|\tilde{\mathbf{x}}\rangle^{\otimes n}\left|\mathbf{x}_{m}\right\rangle^{\otimes n} \pm\left|\mathbf{x}_{m}\right\rangle^{\otimes n}|\tilde{\mathbf{x}}\rangle^{\otimes n}$. In this way, the joint measurement as used above will yield
\begin{eqnarray}
\operatorname{Tr}\left(\sigma_{z}^{(a)} \sigma_{z}^{(l)}\left|\Psi_{f}^{s}\right\rangle\left\langle\Psi_{f}^{s}\right|\right)=\sum_{m=1}^{M}(-1)^{y_{m}} w_{m}\left|\left\langle\tilde{\mathbf{x}} \mid \mathbf{x}_{m}\right\rangle\right|^{2 n}, \nonumber
\end{eqnarray}
and the quantum state fidelity is obtained to determine the result for the classification.
For the preparation of the initial state, it is also mentioned that in many cases creating the desired state can be quite expensive and conditional. 
However, in this work the state preparation does not require the prior knowledge of the training data and the input test data,  and has a relatively low cost. 
All the data can be initially prepared in separate registers in parallel.
Then the Controlled-SWAP gates are applied to the test data register and different training data registers controlled by the index qubits. 
This operation will transform the state 
\begin{eqnarray}
\sum_{m}^{M} \sqrt{w_{m}}|0\rangle_{a}|\tilde{\mathbf{x}}\rangle^{\otimes n}|0\rangle_{d}^{\otimes n}|0\rangle_{l}|m\rangle\bigotimes_{k=1}^{M} |\mathbf{x}_{k}\rangle^{\otimes n}|y_{k}\rangle
\end{eqnarray}
to 
\begin{eqnarray}
\sum_{m}^{M} \sqrt{w_{m}}|0\rangle_{a}|\tilde{\mathbf{x}}\rangle^{\otimes n}\left|\mathbf{x}_{m}\right\rangle_{d}^{\otimes n}\left|y_{m}\right\rangle_{l}|m\rangle \ket{\mathrm{junk_m}}.
\end{eqnarray}
Since the junk register is normalized and will not influence the expectation value $\left\langle\sigma_{z}^{(a)} \sigma_{z}^{(l)}\right\rangle$, the initial state preparation is already accomplished.

Until now, we have introduced several quantum algorithms that utilize kernel methods to handle classification tasks. As mentioned in Ref.
\cite{Schuld2019Quantum},
this approach can be viewed from some different perspectives.
First, the quantum computers can be used to map the classical data $x$ to a Hilbert space state $U_{\phi}(x)|0 \cdots 0\rangle=|\phi(x)\rangle$ and estimate the inner products between these states to obtain the kernel matrix. After that, the rest of the work is left to classical computers by passing the kernel matrix to them.
Indeed, this year a rigorous quantum speedup is proved in Ref. \cite{Liu2021Rigorous},  where a quantum computer is utilized to estimate a kernel function.
In this work, Liu \textit{et al}. constructed a special kind of classification problem which cannot be solved efficiently by a classical computer assuming the widely-believed hardness of the discrete logarithm problem.
Whereas in the supervised learning scenario and with Shor's algorithm \cite{Shor1997PolynomialTime} as a subroutine, the quantum computer is able to efficiently estimate a kernel matrix to map the data to an exponentially large dimensional space and make the training data linearly separable.
In addition to constructing a classification problem in a special situation, this work provides valuable guidance for exploring practical quantum algorithms with advantages over the classical computers.
Second, it is proposed that the classification can be explicitly performed in the feature Hilbert space \cite{Schuld2019Quantum}. For example, the ``linear weight state'' $|w(\theta)\rangle$ can be prepared using a variational quantum circuit $W(\theta)$ such that $|w(\theta)\rangle=W(\theta)|0\rangle$ and the function $f(x , w)=\braket{w|\phi_x}$ can be used as a classification criterion. This idea about using parameterized circuits will be discussed more comprehensively in the following section of variational quantum classifiers.

\subsection{Quantum decision tree classifiers}\label{qdt}

\begin{figure}[htbp]
    \centering
    \includegraphics[width=0.48\textwidth]{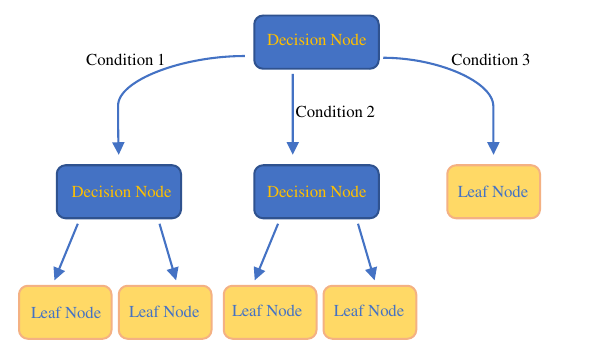}
    \caption{A schematic illustration of a decision tree classifier. Given a new input, it will start at the top node and go down according to its features until arriving at a leaf node. Then the label of the leaf node will be assigned to this input.}
    \label{Fig:DTree}
\end{figure}

A decision tree classifier is a tree-like model and a popular method for classification tasks.
As shown in Fig.\ref{Fig:DTree}, a decision tree classifier is usually built from a top decision node and then split into several branches according to different conditions, which can be viewed as a directed acyclic graph.
Then the leaf nodes at the end of the directed graph represent the classification decisions.
Given a training dataset
$\mathcal{D}$
with $M$ and $d$ being the number of samples and the number of attributes, respectively,
our goal is to build a decision tree according to different features and labels. 
After the building process, given a new unlabeled sample,
we can imagine a point that starts from the top node and heads for a leaf node.
The choice of which way to go at each branch point is decided by the condition that matches the features of the sample.
Finally, the label of the leaf node will be assigned to this sample.

Some basic ideas and algorithms of decision trees have already been summarized in 
\cite{Quinlan1986Induction,Quinlan2014C4.,Breiman1984Classification} including CART algorithm, ID3 algorithm and C4.5 algorithm. 
Due to the decent efficiency and interpretability, 
decision tree method has been already applied in many classification scenarios such as in data mining
\cite{Carvalho2004Hybrid,Kohavi2002Data,Song2015Decision}.

In the first step of building a decision tree, it is important to choose appropriate feature for each node.
Usually, we need to build a subtree using the features that can best improve the classification ability of the current tree,
and the concept of ``information gain'' will be very helpful \cite{Quinlan1986Induction}. For a decision tree denoted by $D$, 
the information gain of $D$ from feature $X$ represents the reduction of the uncertainty of $D$ after knowing the value of $X$, 
which can be formulated as the mutual information:
\begin{eqnarray}
I(D;X)= H(D) - H(D|X),
\end{eqnarray}
where $H(D)$ is the Shannon entropy of $D$ and $H(D|X)$ is the conditional entropy.
After the enumeration of information gain over all features, it is intuitively effective to find the feature that reduces the entropy of the dataset(or sub dataset) most and pick it as the splitting node. This idea of feature choosing is reminiscent of greedy algorithms and is a sub-optimal strategy. We mention that finding the optimal decision tree is an NP-complete problem
\cite{Laurent1976Constructing}.

With the feature choosing strategy introduced above or with some other strategies, e.g. information gain ratio, 
which is a modified version of information gain, 
the decision tree can be build accordingly. 
As the structure of the decision tree may grow complicated during the building procedure, 
it may fail to generalize to classify the unseen data due to overfitting. 
To deal with this problem, a pruning step might be used.
More concretely, 
one can add a penalty term $\alpha |T|$ to the cost function, which can be expressed as 
\begin{eqnarray}
C_{\alpha}(T) = C(T) + \alpha |T|,
\end{eqnarray}
where $|T|$ is the complexity of the tree and $\alpha$ is a tunable hyperparameter.
Then in the pruning process, at each iteration, we will calculate the cost function before and after pruning a leaf node. 
If the cost function becomes smaller after pruning, 
the leaf node will be removed and the algorithm will go on to check other leaf nodes until the procedure is completed.

To build a bridge between decision trees and quantum computation, 
the model of a quantum decision tree classifier has been proposed by Lu and Braunstein
\cite{Lu2014Quantum}. 
Similar to the classical setting, this model starts with a quantum training set
\begin{eqnarray}
\mathcal{D}_{Q}=\left\{\left(\left|x_{1}\right\rangle,\left|y_{1}\right\rangle\right),\left(\left|x_{2}\right\rangle,\left|y_{2}\right\rangle\right), \ldots,\left(\left|x_{M}\right\rangle,\left|y_{M}\right\rangle\right)\right\}. \quad
\end{eqnarray}
Following the classical algorithm, the task for building the decision tree then becomes choosing the attribute $a_i$ that can achieve the best uncertainty reduction. When facing the choice of attributes to split the nodes, the von Neumann entropy is chosen for the quantum case instead of Shannon entropy applied in the classical case.

More recently, In Ref. \cite{Heese2021Representation} the authors have proposed a binary quantum decision tree classifier, named Q-tree, with binary features based on a probabilistic approach.
With a quantum computer, the probabilistic traversal of the decision tree can be accomplished via measurements, and tree inductions and predictions of query data can be integrated into this framework as well.
The quantum circuit model consists of $d + k + m$ qubits and $O(2^{k+m} + d2^d)$ gates, where $d$, $k$, $m$ represent the tree depth, the number of binary features, and the number of binary labels, respectively.
Moreover, this protocol is experimentally demonstrated on a quantum computer via Qiskit using IBM's cloud-based quantum computing service, 
which is the first step to connect the quantum decision tree classifier theory to experimental realizations.

We mention that some other works about quantum decision trees have also been introduced in the literature \cite{Farhi1998Quantum,Shi2002Entropy}, although their major targets may not be classification problems. 
For instance, in Ref. \cite{Farhi1998Quantum}, Farhi and Gutmann proposed quantum decision tree algorithms which are able to solve some computational problems exponentially faster than their classical counterparts.
In addition, in Ref. \cite{Shi2002Entropy} Shi connected the query complexity of quantum decision tree algorithms to Shannon entropy and provided a lower bound for computing any total function $f$, which is an interesting result in the computational complexity field.

\begin{figure*}[htbp]
    \centering
    \includegraphics[width=1\textwidth]{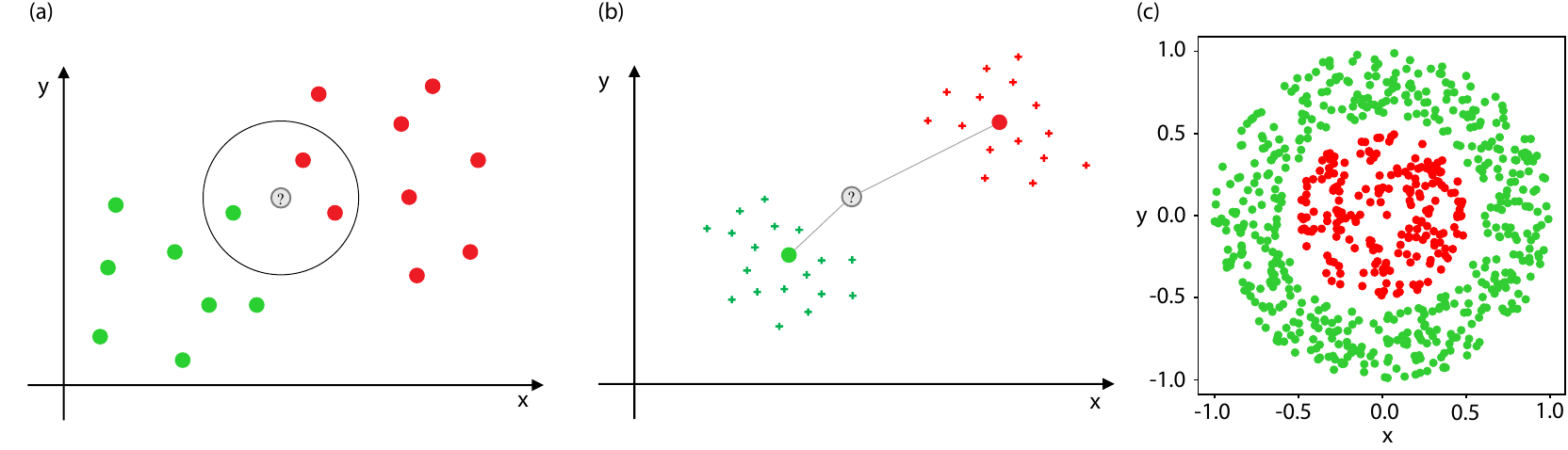}
    \caption{(a) A schematic illustration of the k-nearest neighbors algorithm, where the value of $k$ is three in this case. Here we have an unlabeled data point and two kinds of data points marked as red and green balls, respectively. The label of the majority of the nearest $k$ samples, which is the label of the red balls, will be the output. (b)  A schematic illustration of the nearest centroid algorithm. In this algorithm, the centroids of the different kinds of data points are calculated first. Then the unlabeled data will be assigned to the class corresponding to the closest centroid. (c) A situation that the nearest centroid algorithm may result in poor performance, where the centroid of the two sets lies in the same region.}
    \label{Fig:nearest_neighbor}
\end{figure*}

\subsection{Quantum nearest neighbor algorithm}\label{q nearest objects}

In classical machine learning, the $k$-nearest neighbors algorithm is a non-parametric algorithm which can be applied to classification tasks. 
The original idea was introduced by Cover and Hart
\cite{Cover1967Nearest}.
The basic idea for the $k$-nearest neighbors algorithm is straightforward as indicated by the name. 
Suppose we  have a set of training data,
then with this algorithm we are able to directly use the training data to predict some unseen data's labels without a model training procedure. The classification process is described as follows.

In the first step, we need to define a distance between the feature vectors. 
In the literature, the $L_p$ distance is often chosen as a candidate:
\begin{eqnarray}
L_p(\vec{x}_i,\vec{x}_j) = (\Sigma_{k=1}^N|\vec{x}_{i(k)}-\vec{x}_{j(k)}|^p)^{\frac{1}{p}},
\end{eqnarray}
where $N$ is the dimension of the feature space and $\vec{x}_{i(k)}$ represent the $k$th component of the vector $\vec{x}_i$.
Moreover, the choice of distance measures can also be problem specific. 
For instance, hamming distance can be utilized to handle binary data.
Then for each unseen input to be predicted, 
we search for the nearest $k$ vectors in the training feature set according to the chosen distance measure. 
After obtaining these $k$ vectors with their labels, 
the major label can be taken as the output prediction, 
which is often called the ``majority voting rule''. 
This process is sketched Fig.\ref{Fig:nearest_neighbor}(a).
When applying this strategy, 
the computational cost plays an important role.
If the training set is very large, 
simply calculating all the distances(between the vectors to be predicted and the vectors of the training set) and sorting them will be too computationally expensive. 
To solve this problem, 
sometimes we will need to use some special data structures such as k-d tree
\cite{Bentley1975Multidimensional}
to store the training data and optimize the process of finding the nearest points.
In addition, there are different forms of nearest neighbor algorithms.
As a popular variation, 
the nearest centroid classification is designed to compute the mean value of all samples in the training set labeled $a_i$ to represent the centroid of $a_i$. 
That is, $C_{a_i} = \frac{1}{N_{a_i}}\Sigma_{[y(\vec{x}_{i(k)})=a_i]}\vec{x}_{i(k)}$,
where $N_{a_i}$ is the number of samples labeled $a_i$ and $y$ is the function that outputs the corresponding label.
Then the new data will be categorized to the class according to the nearest centroid, which is illustrated in Fig.\ref{Fig:nearest_neighbor}(b).

After a brief introduction to the classical algorithms, we now introduce two quantum algorithms. First, Lloyd \textit{et al}. have proposed the quantum version of the nearest centroid algorithm
\cite{Lloyd2013Quantum}.
In this task, the key step is to evaluate the distance between the vector $\vec{u}$ to be assigned and the centroids of different labels as shown in Fig.\ref{Fig:nearest_neighbor}(b).
To accomplish this, firstly we write the set of samples of the same label $v$ as $\{|\vec{v_j}|\ket{v_j}\}$ to explicitly show the amplitude and the normalized state.
With the help of quantum random access memory
\cite{Giovannetti2008Quantum},
we prepare the state to be $|\psi\rangle=(1 / \sqrt{2})\left(|0\rangle|u\rangle+(1 / \sqrt{M}) \sum_{j=1}^{M}|j\rangle\left|v_{j}\right\rangle\right)$ in $O(\operatorname{log}(N))$ steps, where $M$ is the size of the subset $\{|\vec{v_j}|\ket{v_j}\}$ and $N$ is the dimension of the vectors.
Then a SWAP test
\cite{Buhrman2001Quantum} can be used to measure the success probability of the first qubit found in $|\phi\rangle=(1 / \sqrt{Z})\left(|\vec{u}||0\rangle-(1 / \sqrt{M}) \sum_{j=1}^{M}\left|\vec{v}_{j}\right||j\rangle\right)$ where $Z$ is a normalization factor. After a simple calculation, the success probability can be written as
\begin{eqnarray}
P = \frac{1}{2Z}\left|\vec{u}-(1 / M) \sum_{j} \vec{v}_{j}\right|^{2}.
\end{eqnarray} 
In other words, we can get an estimation of the target distance by multiplying the success probability by $2Z$. 
The remaining task about constructing the state $\ket{\phi}$ can be accomplished by applying a time evolution of a specially designed Hamiltonian to an easy-to-construct state, which is also discussed in detail in
\cite{Lloyd2013Quantum}.
Finally, by comparing these distances, the classification decision can be made. 
The runtime of the process scales as $O(\operatorname{log}(MN))$, which shows an exponential speedup for both the size of the training set and the dimension of the vectors.

Nevertheless, the classification criterion used above is the distance between the input vector and different centroids. 
In some situations, this might result in a relatively poor performance: when the centroid of the subset lies in the space of other subsets, e.g. the example shown in Fig.\ref{Fig:nearest_neighbor}(c). 
To deal with this issue, Wiebe \textit{et al}. have proposed a quantum nearest-neighbor algorithm to assign a vector $\vec{u}$ to a subset according to the label of $\vec{v_j}$ which satisfies the minimization of $|\vec{u}-\vec{v_j}|$
\cite{Wiebe2014Quantum}.
The main idea can be sketched in three steps.
First,
with the help of quantum random access memory, we can make a query to obtain the state 
$\frac{1}{\sqrt{M}} \sum_{j=1}^{M}|j\rangle\left(\sqrt{1-\left|\vec{v}_{j}-\vec{u}\right|}|0\rangle+\sqrt{\left|\vec{v}_{j}-\vec{u}\right|}|1\rangle\right)$
from
$\frac{1}{\sqrt{M}} \sum_{j=1}^{M}|j\rangle|0\rangle$.
Then the method of amplitude estimation is applied to encode the success probability into an ancillary register
\cite{Brassard2002Quantum}, which can be formulated as
$\frac{1}{\sqrt{M}} \sum_{j=1}^{M}|j\rangle||\vec{v_j}-\vec{u}|\rangle$ with $||\vec{v_j}-\vec{u}|\rangle$ being the quantum state representing the distance $|\vec{v_j}-\vec{u}|$.
In the last step, the Dürr Høyer minimization algorithm
\cite{Durr1996Quantum}, which takes the Grover search algorithm \cite{Grover1997Quantum} as a subroutine, can be applied to search for the closest element to $\vec{u}$ and assign $\vec{u}$ to the corresponding label. This process can be done with a quadratic speedup over the classical counterparts.

\subsection{Quantum annealing based classifiers}

Quantum annealing (QA) is an optimization method for finding the global minimum of a target cost function, especially in situations where many local minima exist \cite{Apolloni1989Quantum,deFalco1988Numerical,Finnila1994Quantum,Kadowaki1998Quantum,Brooke1999Quantum,Crosson2014Different,Farhi2001Quantum,Muthukrishnan2015When}. The feature of this method is well characterized by its name: ``annealing'' originates from annealing in metallurgy---a technique which involves the process of heating and slow cooling.  This provides the basic idea for the simulated and quantum annealing algorithms.

\begin{figure}[htbp]
    \centering
    \includegraphics[width=0.4\textwidth]{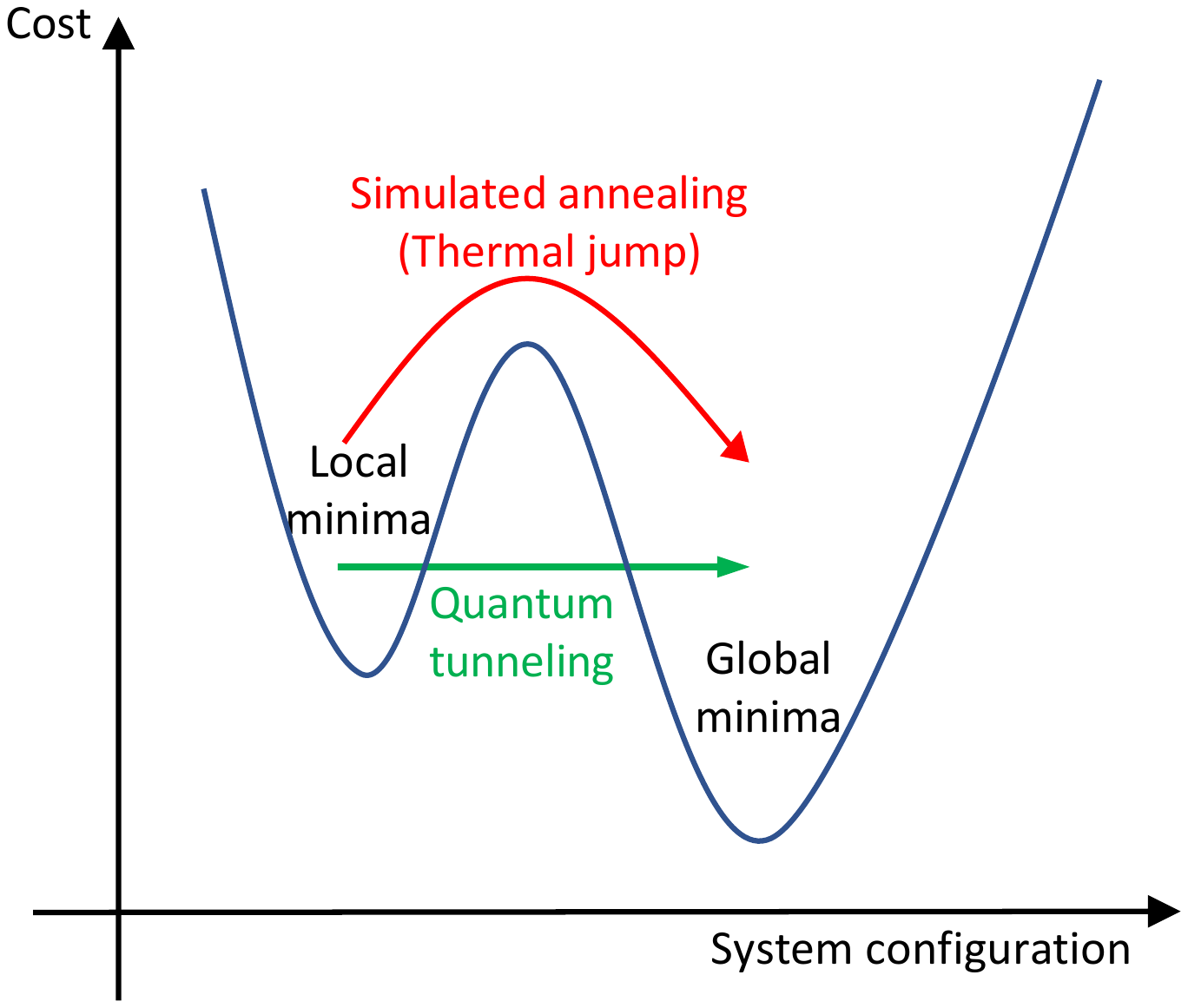}
    \caption{A schematic illustration of quantum annealing. Quantum tunneling is helpful for avoiding the local minima and tunneling through high energy barriers to reach the global solution, while simulated annealing achieves this by thermal jumps.}
    \label{Fig:QA}
\end{figure}

As illustrated in Fig.\ref{Fig:QA}, quantum annealing utilizes quantum tunneling to explore the solution space and find the optimal solution.
The idea is explained as follows.
Consider two Hamiltonians: a ``simple'' one $H_e$ whose ground state is known and easy to prepare, and a `` problem'' one $H_p$ whose ground state corresponds to the solution of a problem.
The system with the ``simple'' Hamiltonian is initialized to the ground state.
Then it is slowly evolved into the ``problem'' Hamiltonian.
During this process, the system is expected to stay in its ground state as ensured by the adiabatic theorem \cite{Born1928Beweis}. The system will end in the ground state of the ``problem'' Hamiltonian and the problem is solved.
Notably, as stated in Ref. \cite{Das2005Quantum}, quantum annealing is able to outperform simulated annealing in certain cases, especially when the cost function landscape consists of high but narrow barriers surrounding shallow local minima.
In 2011, the first commercially available quantum annealer, D-Wave One, was announced by D-Wave Systems \cite{Johnson2011Quantum}, paving the way for developing more advanced devices and solving complex optimization problems \cite{Jones2013Google,Lanting2014Entanglement,Venegas-Andraca2018Crossdisciplinary}.

For the past few years, there are a number of works trying to utilize quantum annealing for classification tasks from different aspects
\cite{Neven2008Training,Neven2009Training,Neven2009Binary,Carapito2021Identification,Li2018Quantum,Li2021Quantumb,Mott2017Solving,Pudenz2013Quantum,Zlokapa2020Quantum,Nath2021Review,Boyda2017Deploying,Caldeira2020Restricted,Denchev2012Robust,Dixit2021Training,DulnyIII2016Developing,Liu2018Adiabatic,Nguyen2018Image,Willsch2020Support}, including simulating quantum annealing based classification algorithms with classical computers and performing classifications with D-Wave quantum annealers. At the early stage of this subfield, Neven \textit{et al.} proposed the training of a binary classifier with a quantum adiabatic algorithm \cite{Neven2008Training}, where the resulting classifier outperforms AdaBoost \cite{Schapire2013Explaining} according to comparisons with a variety of benchmark problems. This protocol has been extended to large scale classification protocols \cite{Neven2009Training} and physical implementations on D-Wave hardware \cite{Neven2009Binary}.
Moreover, a series of works that utilize quantum annealing for different classification tasks have been proposed, including anomaly detection \cite{Pudenz2013Quantum}, Higgs boson classifications \cite{Mott2017Solving,Zlokapa2020Quantum}, classifying and ranking binding affinities \cite{Li2018Quantum}, classifying cancer types and associated molecular subtypes \cite{Li2021Quantumb}, and identifying the driver genes of a severe respiratory response to COVID19 \cite{Carapito2021Identification}, exhibiting notable generalization performance.
In particular, in Ref. \cite{Mott2017Solving}, Mott, Job, Vlimant, Lidar and Spiropulu proposed quantum and simulated annealing methods to solve a Higgs-signal-versus-background machine learning problem, which is mapped to finding the ground state of an Ising spin model.
The classifier consists of a set of weak classifiers based on the kinematic observables of the Higgs decay photons and exhibits robustness against overtraining and errors.
Meanwhile, since the classifiers are functions of experimental parameters that have clear physical meaning, the model provides high interpretability, which is nowadays under active investigations in machine learning \cite{Samek2017Explainable,Zdeborova2020Understanding}.
With the development of quantum annealers, the above introduced algorithms, as well as the quantum annealing based restricted boltzmann machines \cite{Caldeira2020Restricted,Dixit2021Training}, hold intriguing potential in more complex real-life applications.

\subsection{Variational quantum classifiers}\label{qnn}

Artificial neural networks are one of the most powerful approaches in the field of machine learning. 
Since the model of artificial neural networks proposed by McCulloch and Pitts and the concept of perceptron proposed by Rosenblatt in the last century
\cite{Mcculloch1943Logical,Rosenblatt1957Perceptron,Rosenblatt1958Perceptron}, 
this direction has been actively studied and developed for a long time. 
In recent years, 
new breakthroughs of neural networks such as AlexNet \cite{Krizhevsky2017Imagenet} have shown us the magical power of feature extraction and brilliant outlooks for this field.
Now artificial neural networks have already been widely applied in many scenarios from face recognition and natural language processing to robotics, automated driven cars, etc. \cite{Lecun2015Deep,Goodfellow2016Deep}.

Partly inspired by biological neural networks, 
artificial neural networks usually consist of layers of connected artificial neurons and activation functions. 
The parameters (such as weights and biases) that can be optimized during the training procedure are contained in the transformations between the layers.
For an input sample, 
when it goes through the neural networks, 
each layer of the neural networks will make some transformations on it such as linear transformations and pooling operations. 
Then the activation function can be applied to the transformed results to gain more nonlinearity before passing it to the next layer. 

For classification tasks, 
during the training process we have a neural network model where we can input the training data and receive the output result, and the output will indicate the predicted label from the model.
A cost function is often needed to measure the distance between the current output and the training target,
and a proper choice of cost functions will be helpful with the process of optimization.
Some cost functions are commonly used for theoretical analysis and real-life applications, such as the mean square error and cross entropy.
In addition, sometimes in order to improve the generalization behavior and to avoid overfitting,
we will add a regularization term to the cost function, which is relevant to the complexity of the model. With properly chosen loss function $L$, the learning task reduces to an optimization problem \cite{Goodfellow2016Deep}:
\begin{eqnarray}
\min_{f \in \mathcal{F}} \frac{1}{N} \sum_{i=1}^{N} L\left(y_{i}, f\left(\vec{x}_{i}\right)\right)+\lambda R(f),
\end{eqnarray}
where the first term represents the empirical risk and $R(f)$ denotes the complexity of the model with $\lambda$ being a tunable hyperparameter characterizing the importance ratio between them.  
An appropriate choosing of this tunable hyperparameter may greatly improve the training accuracy, while preventing the model from being too complex. This is crucial for the generalization performance of the neural networks.
From another point, 
if two models with different complexity achieve the same empirical error, 
the one with a simpler structure should be taken, 
which is in line with the principle of Occam's razor.
Taking a step forward, with an already well-defined cost function, 
the training process can simply be a process of minimizing the cost function by tuning the weights and biases in the neural networks, 
during which common optimization techniques, such as gradient descent and backpropagation, can be exploited.
After the training process, 
the trained classifiers can be used to make predictions for new unseen data samples. 
Compared with the training process, the computational cost of the predicting process is typically much lower and the trained model can then be encapsulated into some commercial products for real-life applications.

In recent years, quantum neural networks have attracted a lot of  attention and gained rapid development \cite{Adhikary2020Supervised,Cong2019Quantum,Killoran2019Continuous,Grant2018Hierarchical,Schuld2020Circuit,Farhi2018Classification,Blance2021Quantum,Kerenidis2019Quantum,Liu2021Hybrid,Wei2021Quantum,Yano2020Efficient,Li2020Quantuma,Hur2021Quantum,MacCormack2020Branching}.
As a concept originally from the classical machine learning field, 
quantum neural networks usually show up in the form of variational quantum circuits.
Thus, in classification tasks, they are also called variational quantum classifiers.
The variational parameters of these classifiers can be chosen flexibly, e.g. the rotation angle $\theta$ of the component $R_{x}(\theta)$ or some parameterized controlled rotations.
When handling classification tasks, the parameters will be optimized during the training process to minimize a predefined loss function. Then this trained model can be utilized to classify some unseen data.
Based on the above ideas, a variety of structures and protocols have already been proposed. 
To introduce this relatively more diverse subfield, we will split this section into several subsections to make it more organized.
First, we will introduce the gradient descent methods used in the optimization process of variational quantum classifiers. 
Then some popular structures of variational quantum classifiers will be reviewed in the following up subsection. After this, we will turn to introduce some  important phenomena and properties in variational quantum classifiers: the barren plateaus and the vulnerability to adversarial perturbations.

\subsubsection{Gradients evaluation}

\begin{figure}[htbp]
    \centering
    \includegraphics[width=0.48\textwidth]{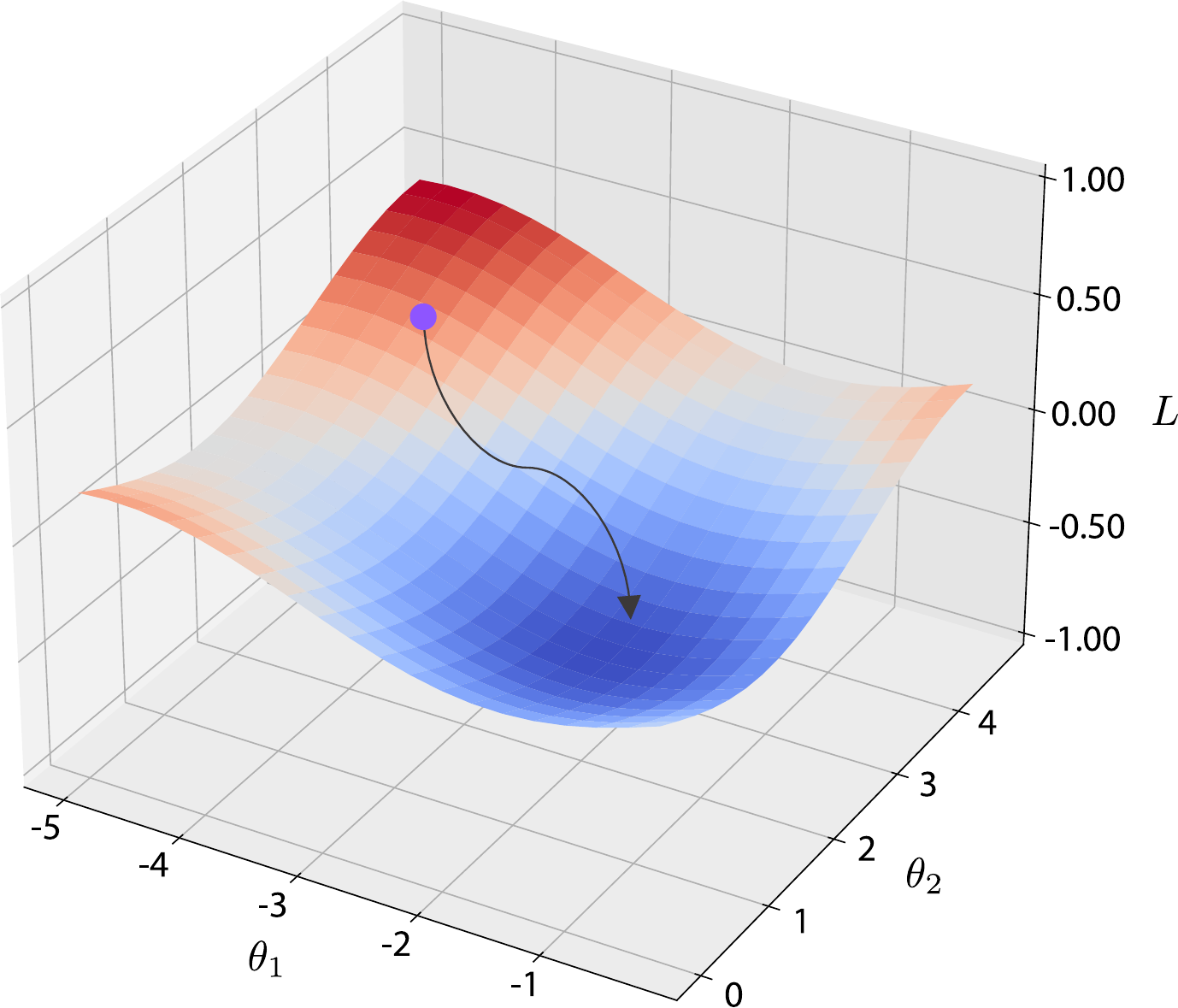}
    \caption{A schematic illustration of the gradient descent process. At each point in the space of the cost function and parameters, the gradient descent strategy provides a direction to effectively reduce the value of the cost function.}
    \label{Fig:gradient}
\end{figure}

When handling classification tasks using neural networks in the classical setting, gradients descent methods and their derivatives are the most frequently used tools during the training process. In general, for a given cost function $L$ with parameters denoted collectively as $\Theta$, the update at gradient descent step $t$ can be expressed as
\begin{eqnarray}
\Theta_{t+1}=\Theta_{t}-\gamma \nabla L\left(\Theta_{t}\right),
\end{eqnarray}
where $\gamma$ is the learning rate. This is an intuitive idea to minimize the cost function, which is illustrated in Fig.\ref{Fig:gradient}.
For practical tasks, with appropriate learning rate and suitable optimizers like Adam \cite{Kingma2014Adam}, the training process can achieve desirable performance.

For this approach, obtaining the gradients usually takes the major computational resources. 
So for a variational quantum circuit, when trying to minimize a cost function, 
it is also wondered whether it is possible to efficiently obtain the gradients during the optimization process. 
Focusing on this question, 
some works have already been done with valuable ideas proposed.
In Refs. \cite{Romero2018Strategies,Li2017Hybrid,Mitarai2018Quantum}, 
it is proposed that if the gate with parameter $\theta$ is in the form $\mathcal{G}(\theta)=e^{-i \theta P_n}$ with $P_n$ being an n-qubit operator in the Pauli group, 
then the derivative can be evaluated with $\frac{\partial\langle B\rangle}{\partial \theta}=\frac{\langle B\rangle^{+}-\langle B\rangle^{-}}{2}$,
where $\langle B\rangle^{\pm}$ means the expectation value of $B$ with the original parameter $\theta$ being $\theta \pm \frac{\pi}{2}$. 
This is usually called the ``parameter shift rule''. 
Compared with the finite difference method expressed as $\frac{\partial\langle B\rangle}{\partial \theta} \approx \frac{\langle B\rangle_{\theta+\frac{\Delta \theta}{2}}-\langle B\rangle_{\theta-\frac{\Delta \theta}{2}}}{\Delta \theta}$, 
the gradient obtained through the parameter shift rule is exact with no discretization error. 
More importantly, 
the result of the exact gradient method is more robust to errors in the near-term quantum devices due to their expressions.
For the sample complexity, since the expectation values are obtained by performing repeated measurements and taking the average value, getting the derivative with respect to a particular parameter with precision $\epsilon$ needs $O(\frac{1}{\epsilon^2})$ measurements according to Chernoff bounds.
Furthermore, 
this ``parameter shift rule'' can be generalized to compute arbitrary order derivatives by applying different shifts to the parameters as proposed in
Ref. \cite{Mari2021Estimating}.

In Refs. \cite{Schuld2020Circuit,Schuld2019Evaluating}, the theory about gradient-based optimization above has been further developed.
In Ref. \cite{Schuld2020Circuit}, the gradient descent scheme is provided for a parameterized circuit with single-qubit gates expressed as 
\begin{eqnarray}
G(\alpha, \beta, \gamma)=\left(\begin{array}{cc}e^{i \beta} \cos (\alpha) & e^{i \gamma} \sin (\alpha) \\ -e^{-i \gamma} \sin (\alpha) & e^{-i \beta} \cos (\alpha)\end{array}\right)
\end{eqnarray}
and their corresponding controlled gates. 
After using analytic methods to express the derivative of the circuit unitary with respect to the parameters, 
the task of computing the partial derivatives can be reduced to measuring the overlap of two quantum states, 
which can be achieved with an inference circuit construction.
In Ref. \cite{Schuld2019Evaluating}, 
the gradient-based approach has been further enriched from several aspects. 
Firstly, the ``parameter shift rule'' gets expanded to a more generalized setting. 
For a gate $\mathcal{G}(\theta)=e^{-i \theta G}$ where G is a Hermitian operator with two distinct eigenvalues $\pm r$, 
the target analytical gradient contains the terms $\frac{1}{\sqrt{2}}(I \pm ir^{-1}G)$ to be constructed. 
Observing the fact that $\mathcal{G}$ can be written in a more convenient form: 
$\mathcal{G}(\theta) = I \cos (r \theta)-i r^{-1} G \sin (r \theta)$,
the above target can be achieved by setting $\theta$ as $\pm \frac{\pi}{4r}$. 
With this idea, 
it is easy to see that the circuit with one-qubit operators generated by Pauli matrices \cite{Mitarai2018Quantum} is a special case for the generalized parameter shift rule. 
Noting that if the operator has more than two eigenvalues, some new strategies need to be developed. The Ref. \cite{Schuld2019Evaluating} has also proposed a possible solution to deal with such a case. 
The key idea is to use an ancillary qubit and decompose the gradient $\partial_{\theta} \mathcal{G}$ into a linear combination of unitary matrices $A_1$ and $A_2$
\cite{Childs2012Hamiltonian}: 
\begin{eqnarray}
\partial_{\theta} \mathcal{G}=\frac{\Gamma}{2}\left[\left(A_{1}+A_{1}^{\dagger}\right)+i\left(A_{2}+A_{2}^{\dagger}\right)\right],
\end{eqnarray}
where $\Gamma$ is a real number depending on the decomposition.  The desired gradient can be evaluated from the probabilities and expectation values, as explained in detail in Ref. \cite{Schuld2019Evaluating}.  The parameter shift rule for Gaussian gates in the continuous variable setting has also been discussed in Ref. \cite{Schuld2019Evaluating}. In addition, we  mention that other methods \cite{Stokes2020Quantum,Koczor2019Quantum, Liu1989Limited,Lavrijsen2020Classical,Haug2021Natural,Wierichs2021General,Sweke2020Stochastic,Koczor2020Quantum,
Guerreschi2017Practical,Kyriienko2021Generalized}, such as quantum natural gradient \cite{Stokes2020Quantum,Koczor2019Quantum} and L-BFGS method \cite{Liu1989Limited,Lavrijsen2020Classical}, have  been introduced  to obtain the gradients as well in the literature. Different methods bear their pros and cons, and the choice of which one to use depends on the specific problem in practice.

\subsubsection{Different structures}
In classical machine learning, different neural networks  are used for different tasks. Various neural networks with different structures have been developed, including feedforward \cite{Bebis1994Feedforward,Svozil1997Introduction}, convolutional \cite{Lawrence1997Face}, recurrent \cite{Mikolov2011Extensions,Zaremba2015Recurrent}, and capsule neural networks \cite{Hinton2011Transforming,Hinton2018Matrix,Sabour2017Dynamic,Xi2017Capsule,Xinyi2018Capsule}.  Similarly, in the emerging field of quantum machine learning, different variational quantum neural networks with distinct structures have also been introduced to tackle  different problems \cite{Killoran2019Continuous,Schuld2020Circuit,Grant2018Hierarchical,Cong2019Quantum,Adhikary2020Supervised,Farhi2018Classification}. In this subsection, we review some important advances about the structures and corresponding optimization strategies of variational quantum classifiers.

\begin{figure}[htbp]
    \centering
    \includegraphics[width=0.48\textwidth]{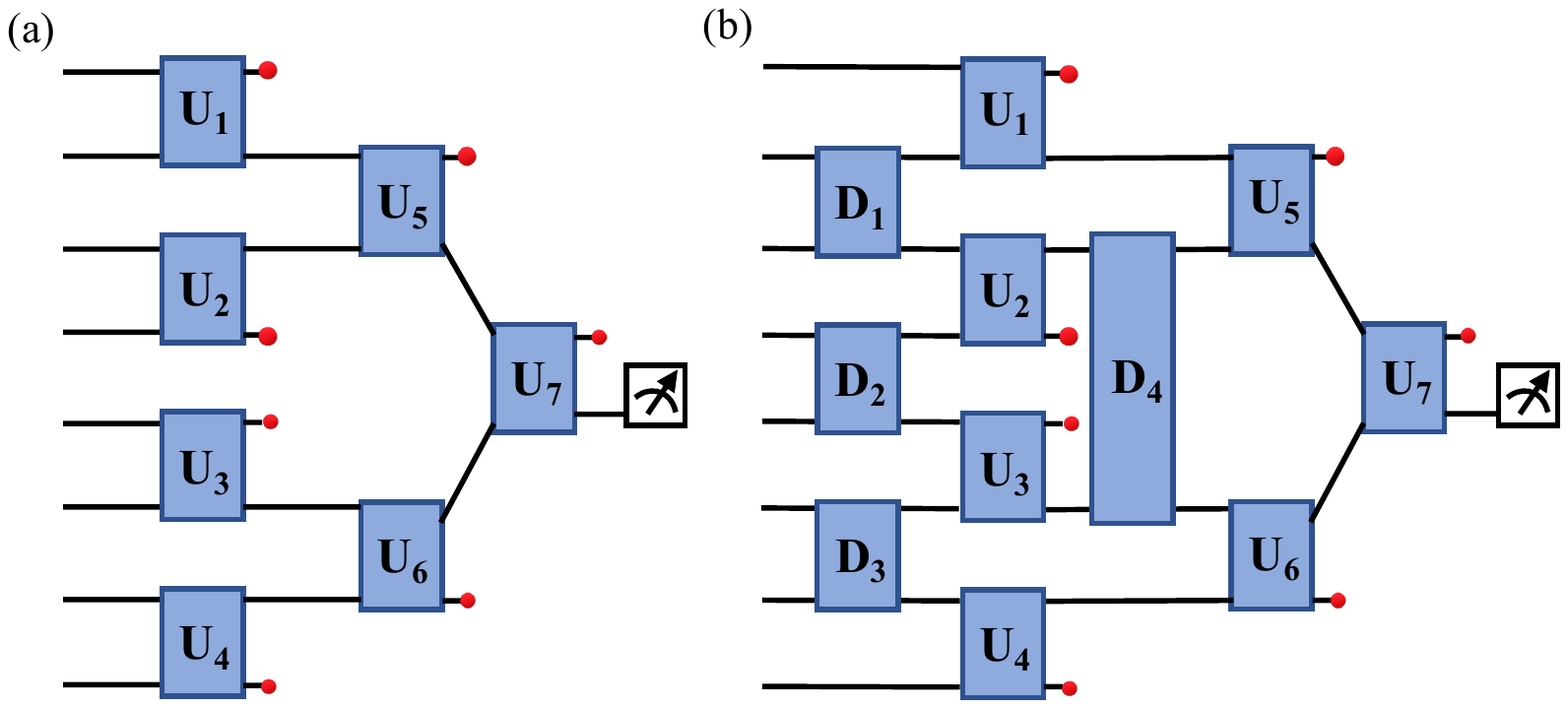}
    \caption{(a) A schematic illustration of a TTN classifier, where a half of the qubits will be traced out after each layer. (b) A schematic illustration of a MERA classifier. The $U_i$ and $D_i$ are unitary blocks.}
    \label{Fig:Hierarchical}
\end{figure}

In Ref. \cite{Grant2018Hierarchical}, a parameterized tree-like quantum classifier has been proposed.
The structures of the tree tensor networks (TTN) \cite{Shi2006Classical} and the multi-scale entanglement renormalization ansatz (MERA) \cite{Vidal2008Class,Cincio2008Multiscale} are chosen as the layouts of the hierarchical circuits as shown in Fig.\ref{Fig:Hierarchical}.
For an input state $\ket{\psi}$, the output is generally an expectation value expressed as
\begin{equation}
M_{\boldsymbol{\theta}}\left(\psi^{d}\right)=\left\langle\psi^{d}\left|\hat{U}_{\mathrm{QC}}^{\dagger}\left(\left\{U_{i}\left(\theta_{i}\right)\right\}\right) \hat{M} \hat{U}_{\mathrm{QC}}\left(\left\{U_{i}\left(\theta_{i}\right)\right\}\right)\right| \psi^{d}\right\rangle, \nonumber
\end{equation}
where $\hat{U}_{\mathrm{QC}}\left(\left\{U_{i}\right\}\right)$ stands for the quantum circuit with parameterized unitaries $U\left(\theta\right)$ and $\hat{M}$ denotes the measurement on a single qubit.
This value can be further used as a component in the optimization and classification process, e.g. as a component in the loss function: $L(\boldsymbol{\theta})=\frac{1}{D} \sum_{d=1}^{D}\left(M_{\boldsymbol{\theta}}\left(\psi^{d}\right)-y^{d}\right)^{2}$, 
where $D$ is the size of the training set and $y^d$ is the label corresponding to the training data of index $d$. 
In this work, different datasets have been used for the training, including Iris dataset, handwritten digits, and quantum data.
For the classical data, 
the qubit encoding method  is chosen instead of amplitude encoding, 
which indicates that the classical data and the quantum state have the same dimension and the element-wise mapping can be written as $\ket{\psi_{n}^{d}}=\cos \left(x_{n}^{d}\right)|0\rangle+\sin \left(x_{n}^{d}\right)|1\rangle$. 
It should be noted that if the dimension of the classical data is beyond the reach of the quantum devices at hand, 
e.g. the $784$ dimensional handwritten digits, dimension reduction methods like principal component analysis (PCA) could be utilized, 
which will inevitably cause some information loss. 
For the quantum data, we typically assume that the data is generated from certain quantum physical process and can be input directly into the quantum classifiers. 
As shown in Ref. \cite{Grant2018Hierarchical}, 
the classifier can achieve decent performance for the above mentioned datasets with proper strategies. 
In addition, the performance of the classifier has been further benchmarked through comparisons between scenarios with and without beforehand PCA dimension reduction for the handwritten digits case, and scenarios with and without ancillary qubits for the quantum data case.

In addition to the classification model above, 
Schuld \textit{et al}. have also proposed a parameterized circuit model with a different structure \cite{Schuld2020Circuit}. 
The circuit is composed of variational single-qubit gates and two-qubit controlled gates. 
Similar to the hierarchical quantum classifiers discussed above, here the training and classification processes make use of the expectation values of the output states as well. 
However, instead of only using this circuit to handle classification tasks on different datasets, 
this work provides several fresh angles to look into variational quantum classifiers.
As already mentioned in the gradient evaluation part, a hybrid gradient descent scheme is given in this work to serve as the subroutine of the training. 
In the numerical simulations, five different datasets are chosen with feature dimensions varying from $13$ to $256$. 
The classification results are listed with benchmarking against six classical models, 
where a comparably good performance is achieved with fewer parameters.
Considering the inevitability of noise in the quantum devices, this work has also numerically studied
the effect of noise in the inputs and circuit parameters, 
and it turns out that this approach is reasonably resilient to noise.

\begin{figure}[htbp]
    \centering
    \includegraphics[width=0.48\textwidth]{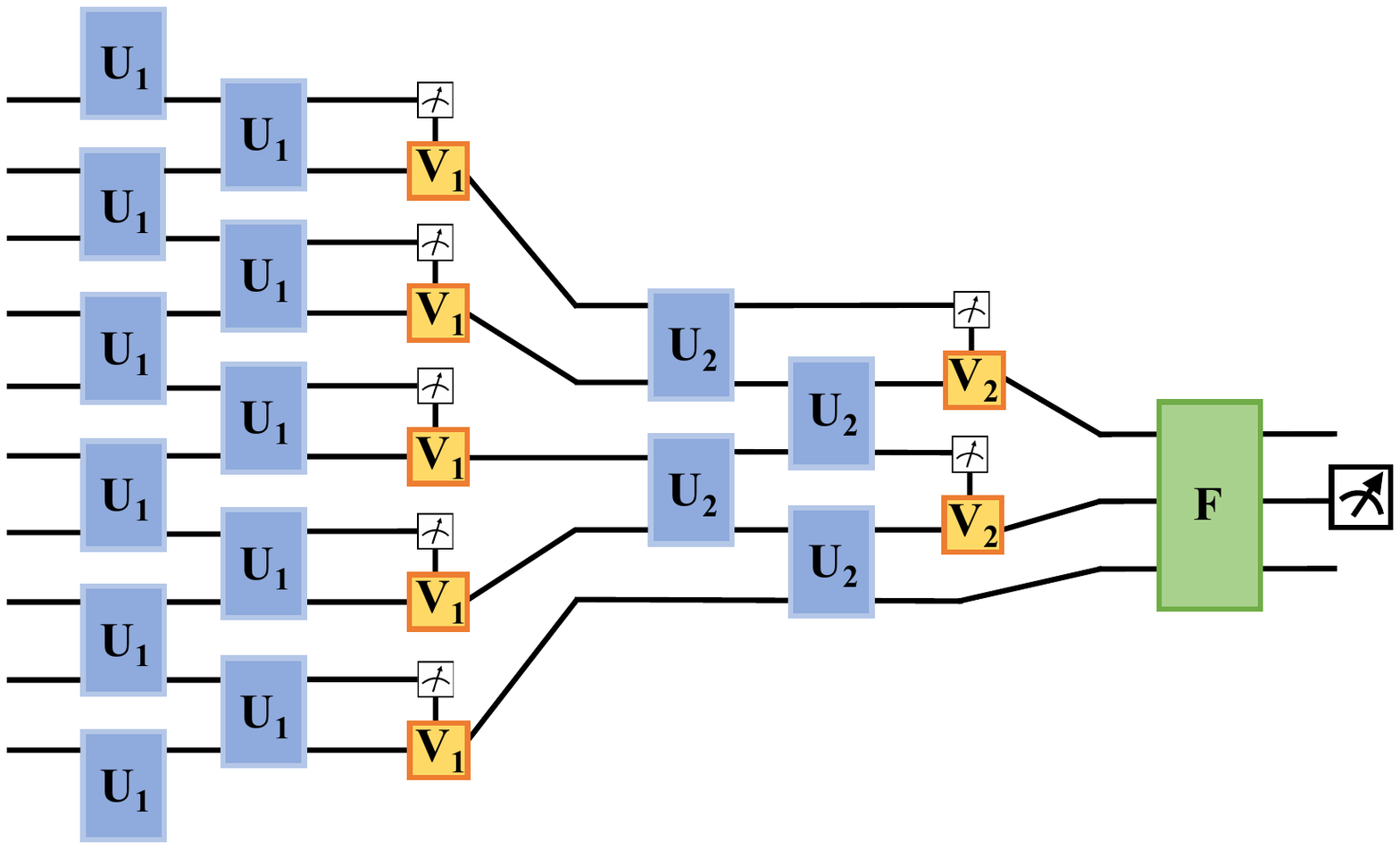}
    \caption{A schematic illustration of the quantum convolutional neural networks, where the $U_i$, $V_i$, and $F$ are unitary blocks and the $V_i$ is controlled by some measurement outcomes.}
    \label{Fig:QCNN}
\end{figure}

Inspired by the classical convolutional neural networks, Cong, Choi, and Lukin introduced a quantum version of convolutional neural networks  in Ref. \cite{Cong2019Quantum}. 
In general, the classical convolutional neural networks consist of  convolution layers, pooling layers and fully connected layers. 
This architecture has been translated into the quantum version as shown in Fig.\ref{Fig:QCNN}.
The convolutional layer in the circuit is represented by the single quasilocal unitary ($U_i$). 
Moreover, the pooling procedure is achieved by measuring a fraction of the qubits and using the outcomes to determine unitary operations ($V_j$) applied to the nearby qubits.
When the number of the remained qubits is reasonably small, a unitary $F$ will serve as a fully connected layer followed by a final measurement. 
During the training process, the unitaries will be optimized with $O(\text{log}(N))$ parameters where $N$ is the number of input qubits. As discussed in Ref. \cite{Cong2019Quantum}, the quantum convolutional neural networks can be  applied to quantum phase recognition to detect a one-dimensional (1D) symmetry-protected topological phase. 
It turns out that the circuit can achieve a high accuracy when the depth is large enough.

For more diverse structures and protocols for variational quantum classifiers, 
there are many other works done from different perspectives \cite{Blance2021Quantum,Kerenidis2019Quantum,Liu2021Hybrid,Wei2021Quantum,Yano2020Efficient,Li2020Quantuma,Hur2021Quantum,MacCormack2020Branching}. 
For example, 
the work in Ref. \cite{Beer2020Training} aims to train a deep quantum neural network to achieve a unitary operation. 
By replacing each target state with a state representing a ``label'', 
it can also be utilized as a classifier.
Moreover, 
the variational quantum classifier can also be built in the continuous-variable architecture
\cite{Killoran2019Continuous}, 
and classical models such as convolutional, recurrent, and residual networks can all be embedded into this setting. 
The variational quantum classifiers can also be utilized in delegated learning protocols, e.g. in Ref. \cite{Li2021Quantum}, the variational classifiers are designed to fit in the blind quantum computation structure \cite{Broadbent2009Universal} as a component in the multiparty federated learning scheme.
Furthermore, there is increasingly more attention paid to the algorithms that search for appropriate structures of the variational quantum circuits
\cite{Rattew2019Domain,Li2017Approximate,Zhang2020Differentiable,Cincio2018Learning,Cincio2021Machine,Lu2020Markovian,Chivilikhin2020Mog,Pirhooshyaran2020Quantum,Ostaszewski2021Structure,Li2020Quantum,Foesel2018Reinforcement,Bolens2020Reinforcement,Wang2021QuantumNAS,Bilkis2021Semiagnostic,Du2020QuantumCircuit,Grimsley2019Adaptive,Tang2021QubitADAPTVQE}, 
which shed light on maximizing the efficiency of quantum devices on the supervised learning tasks. In particular,  a quantum neuroevolution algorithm that autonomously finds near-optimal quantum neural networks for different machine learning tasks has been introduced in Ref. \cite{Lu2020Markovian}. Moreover, in Ref. \cite{Zhang2020Differentiable} the authors have also introduced a differentiable quantum architecture search algorithm, which is capable of automated quantum circuit designs in an end-to-end differentiable fashion.
This field is growing rapidly.
With the development of quantum devices, 
the applications based on these models may bring some practical quantum advantages.

\subsubsection{Barren plateaus}

\begin{figure}[htbp]
    \centering
    \includegraphics[width=0.48\textwidth]{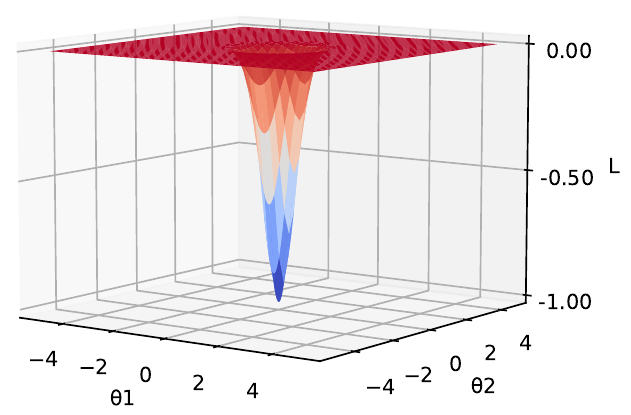}
    \caption{A schematic illustration of the barren plateaus in the optimization process, where  the landscape of the loss function is flat for nearly all the parameter space and the gradient descent methods will fail due to the exponentially vanishing gradients.}
    \label{Fig:BP}
\end{figure}

So far a variety of variational quantum classifiers have been introduced above. 
When a variational quantum circuit is being trained, the update of the parameters in the circuit is expected to efficiently optimize the model.
However, in recent years, several works have shown the existence of gradient vanishing phenomena in the training process of variational quantum circuits.  This is the so-called barren plateau problem, which is illustrated pictorially in Fig.\ref{Fig:BP}. In this subsection, we will review some recent progress on barren plateaus for quantum classifiers. 

In Ref. \cite{Mcclean2018Barren}, McClean \textit{et al}. have shown that, for a wide class of parameterized quantum circuits, the probability that the gradient with respect to any variational parameters is exponentially suppressed as the system size increases. In other words, the loss function landscapes of these variational quantum circuits are mostly flat and barren plateaus  prevail in the training process. To illustrate the essential ideas, we consider the randomly parameterized quantum circuits 
\begin{eqnarray}
U(\boldsymbol{\theta})=U\left(\theta_{1}, \ldots, \theta_{L}\right)=\prod_{l=1}^{L} U_{l}\left(\theta_{l}\right) W_{l},
\end{eqnarray}
where $U_{l}\left(\theta_{l}\right)=\exp \left(-i \theta_{l} V_{l}\right)$ with $V_l$ being a Hermitian operator,  and $W_l$  is a  given unitary without variational parameters. 
The expectation value of the output state over a Hermitian operator $H$ is
$E(\boldsymbol{\theta})=\left\langle 0\left|U(\boldsymbol{\theta})^{\dagger} H U(\boldsymbol{\theta})\right| 0\right\rangle$,
and  the derivative of $E(\boldsymbol{\theta})$ with respect to $\theta_k$ can be expressed as
\begin{eqnarray}
\partial_{k} E \equiv \frac{\partial E(\boldsymbol{\theta})}{\partial \theta_{k}}=i\left\langle 0\left|U_{-}^{\dagger}\left[V_{k}, U_{+}^{\dagger} H U_{+}\right] U_{-}\right| 0\right\rangle,
\end{eqnarray}
where $U_{-} \equiv \prod_{l=0}^{k-1} U_{l}\left(\theta_{l}\right) W_{l}$, $U_{+} \equiv \prod_{l=k}^{L} U_{l}\left(\theta_{l}\right) W_{l}$.
With tools of unitary $t$-designs
\cite{Renes2004Symmetric,Dankert2009Exact,Harrow2009Random},  the authors prove that  if either $U_+$ or $U_-$ is $1$-design, the expectation value of the gradient with respect to the parameter $\theta_k$  is zero, i.e., $\left\langle\partial_{k} E\right\rangle=0$. Moreover, if either $U_+$ or $U_-$ is $2$-design, then an exponential decay of the gradient variance will occur as the system size increases.  
These results explicitly show the exponential concentration of the gradients to zero for the randomly parameterized quantum circuits, which may pose a notable challenge for training quantum classifiers.  From a high-level theoretical perspective, the barren plateaus proved here have a geometric understanding in terms of concentration of measure for high-dimensional spaces \cite{Popescu2006Entanglement,Bremner2009Are,Gross2009Most}.

The presence and absence of barren plateaus depends crucially on the depths of the variational quantum circuits and the properties of the loss functions used. In Ref. \cite{Cerezo2021Cost}, it is proved that if a global loss function is used, then barren plateaus typically appear even for circuits with small constant depths. Whereas, for local loss functions the gradient vanishes at most polynomially as long as the circuit depth is bounded by $O(\log n)$. To obtain an intuitive understanding of these results, the authors considered a toy problem about state preparation with the target state being $|\boldsymbol{0}\rangle$. They started with a tensor-product variational ansatz $V(\boldsymbol{\theta})=\otimes_{j=1}^{n} e^{-i \theta^{j} \sigma_{x}^{(j)} / 2}$, with the goal of finding out the angles $\theta^j$ such that $V(\boldsymbol{\theta})|\boldsymbol{0}\rangle=|\boldsymbol{0}\rangle$.
Assume the global cost function is chosen as
\begin{eqnarray}
C_{G}=\operatorname{Tr}\left[O_{G} V(\boldsymbol{\theta})\left|\psi_{0}\right\rangle\left\langle\psi_{0}\right| V(\boldsymbol{\theta})^{\dagger}\right],
\end{eqnarray}
where $O_{G}=I-|\mathbf{0}\rangle\langle\mathbf{0}|$. This cost function can be simplified as $C_{G}=1-\prod_{j=1}^{n} \cos ^{2} \frac{\theta^{j}}{2}$. 
Direct calculations yield $\left\langle\frac{\partial C_{G}}{\partial \theta^{j}}\right\rangle=0$ and $\operatorname{Var}\left[\frac{\partial C_{G}}{\partial \theta^{j}}\right]=\frac{1}{8}\left(\frac{3}{8}\right)^{n-1}$, which shows that the gradient concentrates exponentially around zero and a barren plateau appears independent of the depth of the circuits. In contrast,  if we choose a local cost function 
\begin{eqnarray}
C_{L}=\operatorname{Tr}\left[O_{L} V(\boldsymbol{\theta})|\boldsymbol{0}\rangle\langle\mathbf{0}| V(\boldsymbol{\theta})^{\dagger}\right],
\end{eqnarray}
where $O_{L}=I-\frac{1}{n} \sum_{j=1}^{n} \ket{0}\bra{0}_{j} \otimes I_{\bar{j}}$ ($I_{\bar{j}}$ denotes the identity operator on all qubits except the one with index $j$),  then direct calculations lead to $C_L = 1-\frac{1}{n} \sum_{j=1}^{n} \cos ^{2} \frac{\theta^{j}}{2}$. In this case, it is straightforward to obtain $\left\langle\frac{\partial C_{L}}{\partial \theta^{j}}\right\rangle=0$ and $\operatorname{Var}\left[\frac{\partial C_{L}}{\partial \theta^{j}}\right]=\frac{1}{8 n^{2}}$,  which indicates that the gradient vanishes polynomially as the system size increases and hence exhibits no barren plateau.

In addition to the above works, we also mention that barren plateaus for variational quantum  circuits have also been investigated from other aspects in the literature \cite{Pesah2020Absence,Wang2020Noise,Huembeli2021Characterizing,Patti2021Entanglement,Marrero2020Entanglement,
Uvarov2021Barren,Arrasmith2020Effect,Wierichs2020Avoiding,Cerezo2021Higher,Grant2019Initialization,
Haug2021Optimal,Holmes2021Barren,Holmes2021Connecting,Zhao2021Analyzing,Kieferova2021Quantum,Liu2021Presence,Skolik2020Layerwise}. For instance,
in Ref. \cite{Pesah2020Absence}, the absence of barren plateaus in quantum convolutional neural networks has been rigorously proved.  This guarantees the trainability of randomly initialized quantum convolutional neural networks and  singles out them as being efficiently trainable unlike many other quantum neural network models. Besides, noise  and entanglement induced barren plateaus have also been investigated in Ref. \cite{Wang2020Noise} and Refs. \cite{Marrero2020Entanglement,Patti2021Entanglement}, respectively.  We also remark that in practical applications the barren plateau problem might be mitigated through a variety of strategies, such as proper initialization \cite{Grant2019Initialization,Verdon2019Learning}, pre-training \cite{Hinton2006Fast,Bengio2007Greedy}, layerwise training \cite{Skolik2020Layerwise}, and judiciously designed quantum circuit structures \cite{Lu2020Markovian,Zhang2020Differentiable}.

\subsubsection{Vulnerability}

The vulnerability of a machine learning model is often relevant to the performance of the model against adversarial attacks.
For a specific example, 
given an input labeled ``panda'' and a fixed classification model,
the unperturbed input will be categorized to the label ``panda'' with high probability.
The adversarial attack aims to generate an imperceptible perturbation on the original input to deceive the classifier, i.e. to efficiently reduce the probability of assigning the input to ``panda'' or to efficiently increase the probability of assigning the input to some wrong labels like ``gibbon''.
This property of machine learning models has already been widely investigated in recent years
\cite{Carlini2017Adversarial,Kurakin2016Adversarial,Miller2019Adversarial,Huang2011Adversarial,
Vorobeychik2018Adversarial,Ilyas2018Black,Tjeng2017Evaluating,Goodfellow2014Explaining,Szegedy2013Intriguing,
Athalye2018Obfuscated,Papernot2017Practical,Madry2017Towards,Biggio2018Wild,Chen2017Zoo}.
Similarly, in the quantum domain, when a quantum classifier is given, it is important to characterize the vulnerability of the model in consideration of adversarial perturbations. In this subsection, we review recent advances about the vulnerability of quantum classifiers.

\begin{figure}[htbp]
    \centering
    \includegraphics[width=0.48\textwidth]{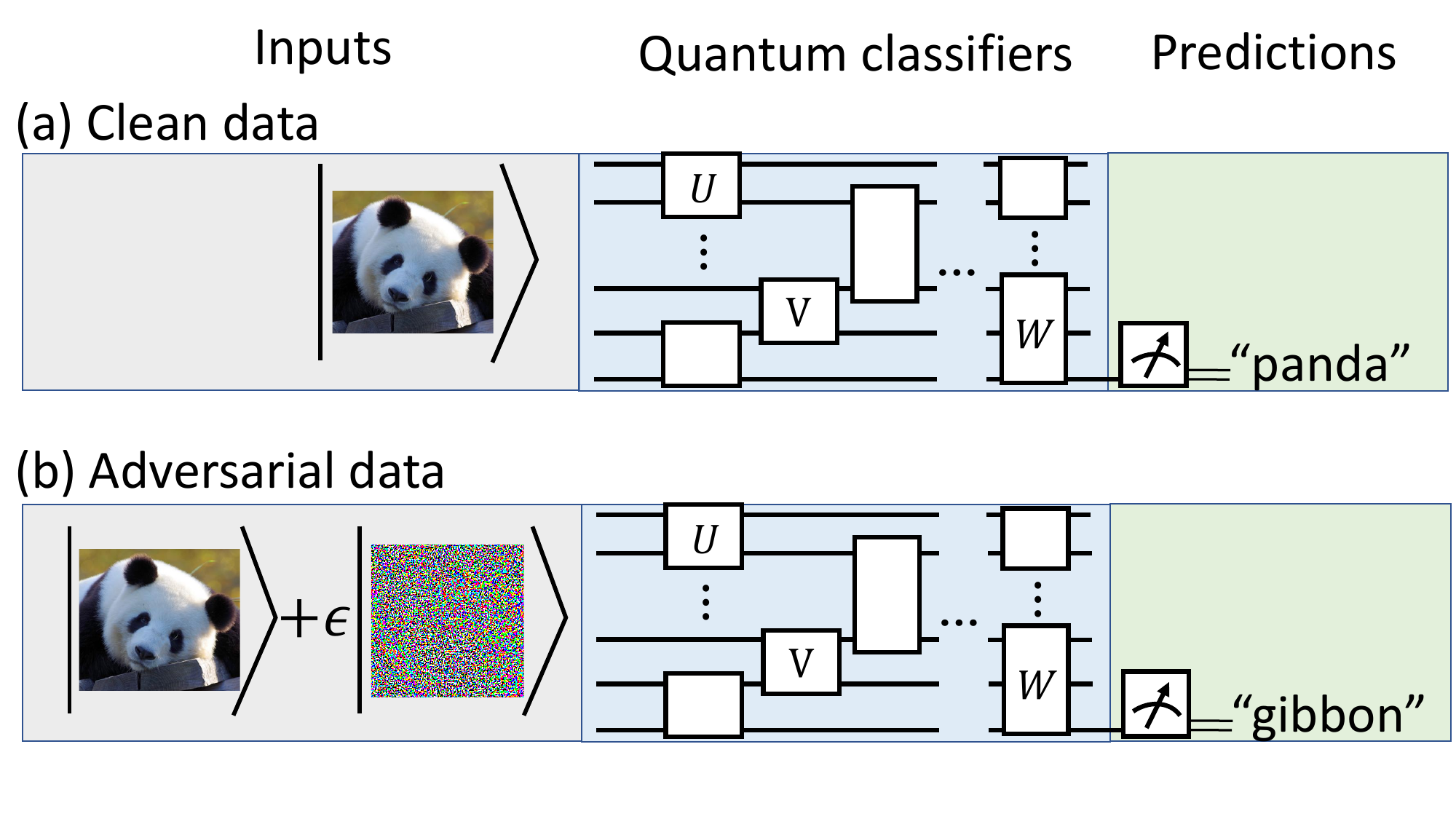}
    \caption{A schematic illustration of quantum adversarial machine learning. (a) The quantum classifier can correctly assign the label ``panda'' to the figure. (b) By adding a tiny amount of carefully-crafted perturbation, the quantum classifier will assign the label ``gibbon'' to the figure incorrectly.}
    \label{Fig:Vulnerability}
\end{figure}

In Ref. \cite{Lu2020Quantum}, Lu, Duan, and Deng studied the vulnerability of quantum machine learning systems in different adversarial scenarios. They found that, similar to classical neural network classifiers, quantum classifiers are likewise vulnerable to crafted adversarial examples, regardless of whether the input data is classical or quantum. In particular, a quantum classifier that achieves nearly the state-of-the-art accuracy can be conclusively deceived by adversarial examples, which are obtained by adding imperceptible perturbations to the original legitimate samples. The essential idea can be illustrated pictorially in Fig.\ref{Fig:Vulnerability}. More concretely, suppose we have a trained model with optimized parameters collectively denoted as $\Theta^*$, the loss function reads $L(h(\ket{\psi}_{\operatorname{in}};\Theta^*),\mathbf{a})$, where $h(\ket{\psi}_{\operatorname{in}};\Theta^*)$ denotes the hypothesis function determined by the variational quantum classifier with parameters fixed to be $\Theta^*$ and $\mathbf{a}$ is the label corresponding to the input state $\ket{\psi}$. An imperceptible perturbation is used as an adversarial attack and implemented as a perturbation operator $U_{\delta}$ acting on $\ket{\psi}_{\operatorname{in}}$.
Here the perturbation operator $U_{\delta}$ can be chosen from a set $\Delta$ of all unitaries close to the identity operator.
For an untargeted attack, 
the goal is to minimize the probability of assigning the right label $\mathbf{a}$ to the state $\ket{\psi}_{\operatorname{in}}$ without caring about which label will be the output. 
So finding out the adversarial perturbations becomes searching a $U_{\delta}$  within $\Delta$ to maximize the loss function:
\begin{eqnarray}
U_{\delta} \equiv \underset{U_{\delta} \in \Delta}{\operatorname{argmax}}\; L\left(h\left(U_{\delta}|\psi\rangle_{\text {in }} ; \Theta^{*}\right), \mathbf{a}\right).
\end{eqnarray}

For a targeted attack, the goal is to maximize the probability to assign a target label $\mathbf{a}^{(t)}$ to the state $\ket{\psi}_{\operatorname{in}}$ where $\mathbf{a} \neq \mathbf{a}^{(t)}$.
In this case, the task can be formalized as finding out the $U_{\delta}^{(t)}$ that minimizes the targeted loss function:
\begin{eqnarray}
U_{\delta}^{(\mathrm{t})} \equiv \underset{U_{\delta}^{(\mathrm{t})} \in \Delta}{\operatorname{argmin}}\; L\left(h\left(U_{\delta}^{(\mathrm{t})}|\psi\rangle_{\mathrm{in}} ; \Theta^{*}\right), \mathbf{a}^{(\mathrm{t})}\right).
\end{eqnarray}
To demonstrate how to obtain adversarial examples in practice, the authors carried out  extensive numerical simulations for several concrete examples covering different scenarios with diverse types of data (such as, handwritten digit images in the MNIST dataset, simulated time-of-flight images in a cold-atom experiment, and quantum data from a one-dimensional transverse field Ising model) and different attack strategies. In addition,  they also discussed  possible defense strategies to enhance the robustness of quantum classifiers against adversarial perturbations. 

On a more analytical level, Liu and Wittek proved rigorously that the amount of perturbation needed for an adversary to induce a misclassification for quantum classifiers scales inversely with dimensionality \cite{Liu2020Vulnerability}. This vulnerability property is a fundamental feature for quantum classifiers independent of the details of the classification protocol, which originates from the concentration of measure phenomenon of high-dimensional Hilbert space.  Furthermore, in Ref. \cite{Gong2021Universal} Gong and Deng studied the universality of adversarial examples and perturbations for quantum classifiers. They proved  that there exist universal adversarial examples that can fool a set of different quantum classifiers: for a set of $k$ classifiers with each input data sample encoded by $n$ qubits, an $O(\frac{\ln k}{2^n})$ increase of the perturbation strength is sufficient to warrant a moderate universal adversarial risk. In addition, they also proved, based on the quantum no free lunch theorem \cite{Poland2020No,Sharma2020Reformulation}, that the universal adversarial risk is  bounded from both below and above and approaches unit exponentially fast as as the size of the classifier increases.

We also mention that there are other works studying quantum adversarial machine learning from different aspects \cite{Du2021Quantum, Liao2021Robust,Weber2021Optimal}. For example, the Ref.  \cite{Du2021Quantum} explored quantum noises to develop a defense strategy against adversarial attacks. Furthermore, in Ref. \cite{Weber2021Optimal} a fundamental link between binary quantum hypothesis testing and probably robust quantum classification has been established, which provides a tight robustness condition for the amount of noise (either natural or adversarial) a classifier can tolerate.

\section{Experimental progress}\label{experiment}
\subsection{Quantum support vector machine and kernel methods}\label{e qsvm}

In the theory section, we have introduced the quantum support vector machine and some other quantum kernel methods.
The extensions from the classical to quantum domain exhibit some attractive properties including potential exponential speedups for certain tasks.
In parallel to the theoretical progress, there are also noteworthy experimental advances and some proof-of-principle physical implementations of these classification models have already been demonstrated in the laboratory \cite{Li2015Experimental,Schuld2017Implementing,Blank2020Quantum,Havlicek2019Supervised,Bartkiewicz2020Experimental}. Here, we  review recent experimental progress on quantum support vector machines and other quantum classification models based on kernel methods.

The first experimental demonstration of quantum support vector machines has been reported in Ref. \cite{Li2015Experimental}.
In this work, the classification of handwritten digits has been experimentally implemented using a small-size quantum circuit. 
Due to limited available qubits, the classification task considered in this experiment was focused on classifying two handwritten digits ``$6$'' and ``$9$''. Moreover, a preprocessing of the images was also carried out to reduce the number of features to $2$:  the vertical ratio and the horizontal ratio according to the pixels in the left (upper) half over the right (lower) half.
The preprocessed images corresponding to digits ``$6$'' and ``$9$'' are defined to be in the positive  and negative classes, respectively.
After these preparations, the quantum support vector machine scheme for this classification task can be delegated to a $4$-qubit nuclear magnetic resonance quantum processor.
As shown in the paper, the experimental predictions  are consistent with the true labels of the handwritten digits.

For the experimental advances in quantum classifiers utilizing kernel methods, several works have been done in recent years
\cite{Schuld2017Implementing,Blank2020Quantum,Havlicek2019Supervised,Bartkiewicz2020Experimental}.
For instance, the distance-based quantum classifier mentioned above
 has been experimentally demonstrated on the superconducting platform of IBM Quantum Experience \cite{Schuld2017Implementing}.
This experiment achieves the classification of the Iris dataset
\cite{Fisher1936Use} using four non-error-corrected superconducting qubits. It is worthwhile to mention that, although the numerical simulation results are close to the theoretical predictions, the experimental results deviate far from them due to the lack of error correction.

In addition, the quantum devices can be used to estimate the kernel functions and the results can be fed to classical computers to do the rest of the training \cite{Schuld2019Quantum,Havlicek2019Supervised,Bartkiewicz2020Experimental,Peters2021Machine,Haug2021Largescale}.
This idea is attractive especially for the stage of noisy intermediate-scale quantum (NISQ) devices, where the quantum technologies are not advanced enough and the cooperation with the classical computers might be helpful.
This hybrid approach has been experimentally demonstrated on both superconducting \cite{Havlicek2019Supervised,Peters2021Machine,Haug2021Largescale} and optical \cite{Bartkiewicz2020Experimental} platforms. 
These results show that classical computers can indeed achieve decent performance by utilizing the kernel matrices evaluated from quantum devices. 
In a more recent work, Ref. \cite{Haug2021Largescale} has utilized the measure of quantum kernels using randomized measurements \cite{Elben2019Statistical,Elben2020CrossPlatform,Zhu2021CrossPlatform}, which is capable of handling large datasets with a quadratic speedup. Moreover, the data in this work is encoded into parameterized quantum circuits, and the expressive power of the encoding strategy has a theoretical guarantee \cite{Haug2021Capacity}. After the kernel estimation, the kernel can be further processed by classical SVMs. This setting has been successfully applied to demonstrate the classification of the $36$ and $64$ dimensional handwritten digit datasets on the IBM quantum computer with error mitigation techniques, while other methods such as the swap test \cite{Buhrman2001Quantum} and the inversion test \cite{Havlicek2019Supervised} may take much longer time due to the quadratic scaling with the dataset size.

\subsection{Quantum nearest neighbor algorithm}\label{q_knn}

As mentioned in the above sections, with quantum random access memory, the quantum nearest neighbor algorithm has the potential to offer an exponential speedup. Moreover, the first proof-of-principle demonstration has also been implemented \cite{Cai2015Entanglement}.
This experiment is conducted on the optical platform to assign different entangled states to their nearest clusters.
Here, the two clusters $A$ and $B$ are represented by reference vectors $\vec{v}_A$ and $\vec{v}_B$, respectively.
For the job of assigning a new vector $\vec{u}$, it can be transformed into comparing the distance between $\vec{u}$ and the reference vectors: $D_{A}=\left|\vec{u}-\vec{v}_{A}\right|, \text { and } D_{B}=\left|\vec{u}-\vec{v}_{B}\right|$ and assigning it to the cluster with a smaller distance.

To achieve this goal, the single photons are used as qubits and $\ket{0}$ and $\ket{1}$ are encoded with their horizontal and vertical polarization.
A key step is to create the entanglement resource states for data encoding.
Firstly, the polarization-entangled photon pairs $\left(|0\rangle_{\mathrm{anc}}|0\rangle_{\mathrm{vec}}+|1\rangle_{\mathrm{anc}}|1\rangle_{\mathrm{vec}}\right) / \sqrt{2}$, where one photon is used as an ancillary qubit, can be generated through spontaneous parametric down-conversion
\cite{Kwiat1995New}.
Then the four-photon resource states can be generated using two entangled photon pairs with a post-selection scheme, which results in the four-qubit Greenberger-Horne-Zeilinger entangled state
\cite{Pan2012Multiphoton}:
\begin{eqnarray}
\frac{1}{\sqrt{2}}\left(|0\rangle_{\mathrm{anc}}|000\rangle_{\mathrm{vec}}+|1\rangle_{\mathrm{anc}}|111\rangle_{\mathrm{vec}}\right).
\end{eqnarray}
The three-qubit state can be generated straightforwardly by a projection.

With these states prepared, the data encoding can be accomplished by firstly sending the single photons through a polarizing beam splitter and splitting them into two spatial modes. Then the controlled unitary gates can be applied to these two spatial modes, respectively
\cite{Zhou2013Calculating}. After recombining these two modes on a nonpolarizing beam splitter, the final state $\left(|0\rangle\left|u_{1}\right\rangle_{\text {new }}+|1\rangle\left|v_{1}\right\rangle_{\text {ref }}\right) / \sqrt{2}$ can be generated. With all the preparations, the classification scheme can be deployed on the optical platform. It turns out to achieve decent classification accuracy of the two-, four-, and eight-dimensional vectors that are encoded in the two-, three-, and four-qubit entanglement states. The misclassified cases partly come from ``boundary conditions'', where the error is comparable to the distance $|D_A-D_B|$ and the performance of the distance evaluation which is affected by the fidelity of the entangled states \cite{Cai2015Entanglement}.

More recently, a nearest centroid classifier has been experimentally demonstrated on a eleven-qubit trapped ion quantum computer \cite{Johri2021Nearest}, and the quantum device is now commercially available via IonQ’s cloud service. In this work, the authors proposed a quantum circuit to help estimate the Euclidean distance between two $d$-dimensional data points $x$ and $y$:
\begin{eqnarray}
I_{x y}=\sqrt{\|x\|^{2}+\|y\|^{2}-2\|x\|\|y\| c_{x y}},
\end{eqnarray}
where $c_{x y}=\langle x \mid y\rangle$ is the inner product of the two normalized vectors.
In this protocol, estimating the distance between $k$ centroids and $n$ $d$-dimensional data points will need a $d$-qubit quantum circuit with running time $O(kd+nd+kn\operatorname{log}(d)/\epsilon)$.
This framework is experimentally applied to the classification of a synthetic dataset and a PCA-processed MNIST handwritten digit dataset, thus providing beneficial benchmarks for future works.

\subsection{Variational quantum classifiers}\label{vqc_e}

Variational quantum classifiers have drawn broad interest over recent years. However, due to the system size limitation of current quantum devices, most of the classifiers reviewed in the above sections are not readily feasible to be implemented with a large number of qubits in the laboratory at the current stage. To connect the theories with the experimental implementations, some proof-of-principle experiments have been carried out. In this subsection, we will introduce several experimental advances of the classifiers composed of variational quantum circuits
\cite{Grant2018Hierarchical,Havlicek2019Supervised}.

\begin{figure}[htbp]
    \centering
    \includegraphics[width=0.48\textwidth]{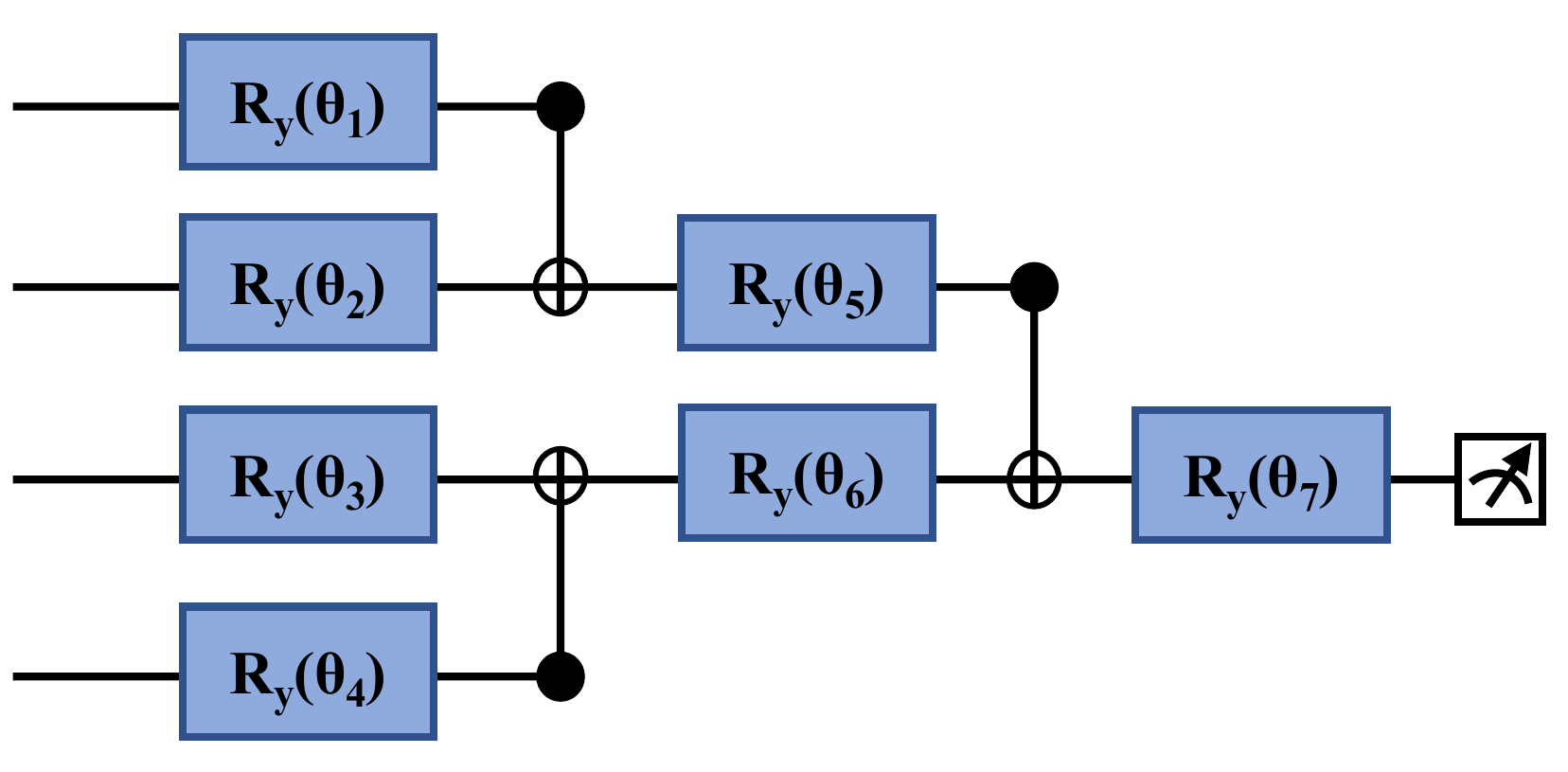}
    \caption{The structure of a proof-of-principle TTN classifier.}
    \label{Fig:expe_Hierarchical}
\end{figure}

In the above sections, we have introduced the structures of hierarchical quantum classifiers including tree tensor networks and multi-scale entanglement renormalization ansatz \cite{Grant2018Hierarchical}.
In Ref. \cite{Grant2018Hierarchical}, a proof-of-principle experiment has been demonstrated to classify two classes of the Iris data, where a tree tensor network quantum classifier is deployed on the ibmqx4 quantum computer.
The circuit is designed as shown in Fig.\ref{Fig:expe_Hierarchical} and it is trained classically first and then deployed on the quantum device, which is finally able to implement the classification of the test data with $100\%$ accuracy.

\begin{figure}[htbp]
    \centering
    \includegraphics[width=0.48\textwidth]{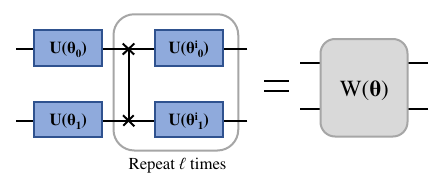}
    \caption{The structure of the variational quantum circuit for the training task. The entangling gates used in this protocol are controlled-Z gates and the local gates belong to SU(2).}
    \label{Fig:expe_Classification}
\end{figure}

In Ref. \cite{Havlicek2019Supervised}, a variational quantum classifier has been experimentally implemented on the superconducting platform to classify artificially generated data.
To start with, the training data $\boldsymbol{x}$ is encoded in a quantum state $\ket{\Phi(\boldsymbol{x})}$ by applying an operator $\mathcal{U}_{\Phi(\boldsymbol{x})}=U_{\Phi(\boldsymbol{x})} H^{\otimes n} U_{\Phi(\boldsymbol{x})} H^{\otimes n}$ to $\ket{0}^n$ where the component $U_{\Phi(\boldsymbol{x})}$ can be expressed as   \cite{Havlicek2019Supervised}:
\begin{eqnarray}
U_{\Phi(\boldsymbol{x})}=\exp \left(i \sum_{S \subseteq[n]} \phi_{S}(\boldsymbol{x}) \prod_{i \in S} Z_{i}\right),
\end{eqnarray}
with $\phi_{\{i\}}(\boldsymbol{x})=x_{i}$ and $\phi_{\{1,2\}}(\boldsymbol{x})=\left(\pi-x_{1}\right)\left(\pi-x_{2}\right)$.
Then to assign the training data into different classes, 
a random unitary $V \in S U(4)$ and the observable $\boldsymbol{f}$=$Z_{1} Z_{2}$ are chosen. 
The label of $\boldsymbol{x}$ will be assigned to $+1$ if $\left\langle\Phi(\boldsymbol{x})\left|V^{\dagger} \boldsymbol{f} V\right| \Phi(\boldsymbol{x})\right\rangle \geq \Delta$ and to $-1$ if $\left\langle\Phi(\boldsymbol{x})\left|V^{\dagger} \boldsymbol{f} V\right| \Phi(\boldsymbol{x})\right\rangle \leq-\Delta$ with $\Delta$ being $0.3$. 
After these data preparations, 
the short-depth variational circuit $W(\theta)$ designed as shown in Fig.\ref{Fig:expe_Classification} can be applied to the data states and takes the role of assigning the data to different categories. 
For the basic settings of the training process, 
the binary measurement $\{M_y\}$ in the $Z$-basis is performed on the state $W(\theta) \mathcal{U}_{\Phi(\boldsymbol{x})}|0\rangle^{n}$ with success probabilities $p_{y}(\boldsymbol{x})=\left\langle\Phi(\boldsymbol{x})\left|W^{\dagger}(\boldsymbol{\theta}) M_{y} W(\boldsymbol{\theta})\right| \Phi(\boldsymbol{x})\right\rangle$. 
In the physical experiments, 
these probabilities can be evaluated by repeating the measurement and taking the empirical distribution $\hat{p}_y$. 
Then the data $\boldsymbol{x}$ will be assigned to class $y$ if $\hat{p}_{y}(x)>\hat{p}_{-y}(x)-y b$, 
where the bias parameter $b \in [-1,1]$ can also be optimized. 
During the training process, 
the cost function is defined as the error probability and it is optimized with Spall's simultaneous perturbation stochastic approximation algorithm
\cite{Spall1997One,Spall2000Adaptive}.
After the training procedure, the parameters of the circuit are fixed and can be used for classifying some unseen data. 
With these settings, the experiments with error mitigation techniques and different circuit depths have been carried out, 
where a $100\%$ success rate can be achieved with the depth four circuit.

\section{Conclusion and outlooks}

Over the past two decades,
the world has witnessed revolutionary development in both quantum computing and machine learning.
As an appealing interdisciplinary application, the quantum classifiers have attracted a wide range of interest.
The models of quantum classifiers have been developed from a number of aspects.
Some quantum classifiers can be viewed as the quantum extensions of some classical classification algorithms.
What is more, these quantum classifiers may have the potential to exhibit advantages over their classical counterparts.
In addition,
there are other quantum classifiers that utilize the structures  drawing inspiration from physical concepts,
e.g. the tree tensor network classifiers, the multi-scale entanglement renormalization ansatz classifiers, and the multi-level quantum system classifiers.
These explorations might be helpful to find models that are feasible to tackle certain classification problems with some inherent physical structures.

Yet, despite the rapid development, there are crucial challenges for quantum classifiers that stand in the way to future practical applications.
First, for the quantum devices,
the quantum random access memory is a key component for a number of algorithms to demonstrate the potential advantages,
while it is still far from implementing QRAM with a desirable scale at the current stage.
Second, since the noise in the quantum devices seems inevitable,
increasing the fidelity of the quantum operations and developing better error correction techniques are of crucial importance.
Third, for the variational quantum classifiers,
there are already some works trying to measure their representation power \cite{Abbas2021Power,Meyer2021Fisher,Wang2021Understanding,Schuld2021Effect,Caro2021Encoding,Wu2021Expressivity,Funcke2021Dimensional,Banchi2021Generalization,Du2021Efficient,Du2020Learnability,Haug2021Capacity}.
To find whether there is a separation between these models and their classical counterparts for potential future applications, further studies are highly desirable.
In addition, the barren plateau phenomena and the vulnerability of quantum classifiers to adversarial attacks mentioned in this review also present the potential problems against the future applications, though there are already a variety of works trying to formalize and handle them.
For the future development of this field, it is important to further develop quantum classification theories and models to find more potential applications. 
With the help of the softwares used for simulating these quantum models \cite{Luo2019Yao.,Bezanson2017Julia,Broughton2020TensorFlow,Bergholm2020PennyLanea,Killoran2019Strawberry,Aleksandrowicz2019Qiskit,Svore2018Enabling,Zhang2019Alibaba,Huang2019Alibaba,Nguyen2021HiQProjectQ,Green2013Quipper,JavadiAbhari2015ScaffCC,Khammassi2017QX,Johansson2012QuTiP}, it gets easier to quantify the performance of these models.
Moreover, the size of the quantum platforms, such as the superconducting platforms and the ion trap systems, are expected to grow fast to meet the requirement of the theories of the various quantum classifiers.
The combination of the advanced theories of quantum classifiers and the moderate-size quantum computers may demonstrate useful advantages in practical applications in the near future.

From the perspective of computational complexity, in Ref. \cite{Bittel2021Training} Bittel and Kliesch proved that the classical optimization problems for variational quantum algorithms are NP-hard, even for classically tractable systems which are composed of only logarithmically many qubits or free fermions. This work showed the intrinsic hardness for the classical optimization and the training landscape may have many local minima far from optimal, which presents a severe challenge for variational quantum classifiers.
In addition, the quantum advantage in machine learning can also be viewed in terms of the number of times to access a quantum process $\mathcal{E}$ in the classical or quantum setting. In Ref. \cite{Huang2021InformationTheoretic}, Huang, Kueng and Preskill proved that for any input distribution, a classical machine learning model can achieve accurate predictions \textit{on average} by accessing $\mathcal{E}$ with times comparable to the optimal quantum machine learning model. By comparison, an exponential quantum advantage is possible for achieving an accurate prediction \textit{on all inputs}.
For instance, predicting the expectations of all Pauli observables in an $n$-qubit system $\rho$ requires $O(n)$ copies of $\rho$ in a quantum machine learning model, exponentially less than $2^{\Omega(n)}$ copies in a classical one. 
Moreover, the sample complexity has also been investigated in Ref. \cite{Bu2021Statistical,Cai2021Sample}.
These works provide helpful insights for understanding quantum advantages in machine learning tasks.

Without a doubt, it is hard to say when the first quantum classifier will be commercially available. Harder even is to predict what special advantage the first commercial quantum classifier will bring us in real life. 
This emergent research direction is largely still in its infancy, with many important issues remaining barely explored.
Yet, one thing is for sure: the combination of supervised learning and quantum physics is a win-win cooperation that has great potential to revolutionize many aspects of our modern world.

We would like to thank Zhide Lu, Wenjie Jiang, Daniel Lidar, Tobias Haug, Yuxuan Du, Peixin Shen, Sirui Lu, and Liwei Yu for helpful discussions and communications. This work is supported by the start-up fund from Tsinghua University (Grant. No. 53330300320), the National Natural Science Foundation of China (Grant. No. 12075128), and the Shanghai Qi Zhi Institute.

\bibliography{QMLbib}
\end{document}